\pdfoutput=1
\documentclass{ws-ijmpa}
\usepackage[super,compress]{cite}
\usepackage{graphicx}
\usepackage{xspace}
\usepackage{doi}
\usepackage{subfigure}
\usepackage{floatrow}
\usepackage{textcomp}
\usepackage{subscript}

\begin{document}
\markboth{C.~Alduino et al.}{}
\def\qz{\mbox{CUORE-0}\xspace}
\def\qino{\mbox{CUORICINO}\xspace}

\def\mcuorez{\emph{MCuoreZ}\xspace} 
\def\pbext{\emph{ExtPbS}\xspace} 
\def\cryoext{\emph{CryoExt}\xspace} 
\def\cryoint{\emph{CryoInt}\xspace} 
\def\pbint{\emph{IntPb}\xspace} 
\def\holder{\emph{Holder}\xspace}
\def\smallpart{\emph{Small Parts}\xspace}
\def\crystal{\emph{Crystals}\xspace}



\def\mspec{$\mathcal{M}$\xspace}
\def\mspecone{$\mathcal{M}_1$\xspace}
\def\mspectwo{$\mathcal{M}_2$\xspace}
\def\mspecthree{$\mathcal{M}_3$\xspace}
\def\summspec{$\Sigma$\xspace}
\def\summspectwo{$\Sigma_2$\xspace}


\def\u{$^{238}$U\xspace}
\def\ths{$^{232}$Th\xspace}
\def\thdto{$^{230}$Th\xspace}
\def\acddo{$^{228}$Ac\xspace}
\def\raddq{$^{224}$Ra\xspace}
\def\rndd{$^{222}$Rn\xspace}
\def\poduq{$^{214}$Po\xspace}
\def\bidq{$^{214}$Bi\xspace}
\def\pbduq{$^{214}$Pb\xspace}
\def\podud{$^{212}$Po\xspace}
\def\bidud{$^{212}$Bi\xspace}
\def\pbdud{$^{212}$Pb\xspace}
\def\podd{$^{210}$Po\xspace}
\def\pbdd{$^{210}$Pb\xspace}
\def\tld{$^{208}$Tl\xspace}
\def\bidzs{$^{207}$Bi\xspace}
\def\pt{$^{190}$Pt\xspace}
\def\euucq{$^{154}$Eu\xspace}
\def\euucd{$^{152}$Eu\xspace}
\def\csuts{$^{137}$Cs\xspace}
\def\xeuts{$^{136}$Xe\xspace}
\def\xect{$^{130}$Xe\xspace}
\def\tect{$^{130}$Te\xspace}
\def\tecv{$^{128}$Te\xspace}
\def\tecw{$^{120}$Te\xspace}
\def\tinw{$^{120}$Sn\xspace}
\def\teudcm{$^{125m}$Te\xspace}
\def\sbudc{$^{125}$Sb\xspace}
\def\tecvt{$^{123}$Te\xspace}
\def\cduus{$^{116}$Cd\xspace}
\def\cdcd{$^{110}$Cd\xspace}
\def\agcd{$^{110}$Ag\xspace}
\def\aguzom{$^{108m}$Ag\xspace}
\def\rhuzdm{$^{102m}$Rh\xspace}
\def\rhuzu{$^{101}$Rh\xspace}
\def\mouzz{$^{100}$Mo\xspace}
\def\seod{$^{82}$Se\xspace}
\def\gect{$^{76}$Ge\xspace}
\def\cosz{$^{60}$Co\xspace}
\def\cocs{$^{57}$Co\xspace}
\def\mncq{$^{54}$Mn\xspace}
\def\caqo{$^{48}$Ca\xspace}
\def\kq{$^{40}$K\xspace}

\def\teod{TeO$_2$\xspace}

\def\camo{CaMoO$_4$}
\def\camoenr{$^{40}$Ca$^{100}$MoO$_4$}
\def\camodepl{$^{40}$Ca$^{\mathrm{nat}}$MoO$_4$}
\def\cdwo{CdWO$_4$}
\def\caf{CaF$_2$}


\def\gm{$\gamma$\xspace}
\def\gagamma{g_{a\gamma\gamma}\xspace}
\def\alph{$\alpha$\xspace}
\def\bmin{$\beta^-$\xspace}

\def\bb{$\beta\beta$\xspace}
\def\bbd{$2\nu\beta\beta$\xspace}
\def\bbz{$0\nu\beta\beta$\xspace}
\def\bmbm{$\beta^-\beta^-$\xspace}
\def\bpbp{$\beta^+\beta^+$\xspace}
\def\bpEC{EC$\beta^+$\xspace}
\def\ECEC{EC-EC\xspace}

\def\mbb{$|\langle m_{\beta \beta}\rangle |$}
\def\me{$m_e$}
\def\Qbb{$Q_{\beta\beta}$}
\def\FD{${F}_{D}^{0\nu}$}
\def\FZB{$F_{ZB}^{0\nu}$}
\def\Tzn{$T_{1/2}^{0\nu}$}
\def\Tdn{$T_{1/2}^{2\nu}$}
\def\rate{counts/(keV$\cdot$kg$\cdot$y)}
\def\vita-dim{$\tau_{1/2}$}

\def\QE{$\epsilon_Q$}
\def\ENCe{$\mathtt{ENC_e}$}
\def\ENC{$\mathtt{ENC}$}
\def\bg{$\beta/\gamma$}

\def\avl{$\langle \lambda \rangle$~}
\def\ave{$\langle \eta \rangle$~}
\def\avm{$\langle g_{\chi\nu} \rangle$~}
\def\mee{$\langle m_{ee} \rangle$~}
\def\mnu{$\langle m_{\nu} \rangle$~}
\def\amnu{$\vert\langle m_{\nu} \rangle\vert$\xspace}
\def\nmamnu{\vert\langle m_{\nu} \rangle\vert}
\def\mmod{$\| \langle m_{ee} \rangle \|$}
\def\mb{$\langle m_{\beta} \rangle$~}
\def\fn{$G^{0\nu}\vert M^{0\nu} \vert ^2$}
\def\nmfn{G^{0\nu}\vert M^{0\nu} \vert^2}
\def\Mz{$|M_{0\nu}|$~}
\def\Md{$|M_{2\nu}|$~}
\def\Tz{$T^{0\nu}_{1/2}$~}
\def\Td{$T^{2\nu}_{1/2}$~}
\def\Tm{$T^{0\nu\,\chi}_{1/2}$~}
\def\ca{$\sim$}
\def\dca{$\approx$}
\def\dot{$\cdot$}
\def\pom{$\pm$ }

\def\ne{$\neq$}
\def\be{\begin{equation}}
\def\ee{\end{equation}}
\def\gohm{G$\Omega$}
\def\ohm{$\Omega$}
\def\per{$\times$}
\def\less{$<$}
\def\ciccio{5$\times$5$\times$5 cm$^3$ }
\def\magro{3$\times$3$\times$6 cm$^3$ }
\def\rate{counts/(keV$\cdot$kg$\cdot$y)}
\def\rateEnv{counts$\cdot$10$^{-3}$)/(kg$\cdot$keV$\cdot$y}

\def\ums{$\mu$m}

%
\catchline{}{}{}{}{}
%

\title{Study of Rare Nuclear Processes with CUORE}

\maketitle
\noindent
C.~Alduino$^{1}$, K.~Alfonso$^{2}$, F.~T.~Avignone~III$^{1}$, O.~Azzolini$^{3}$, G.~Bari$^{4}$, F.~Bellini$^{5,6}$, G.~Benato$^{7}$, A.~Bersani$^{8}$, M.~Biassoni$^{9}$, A.~Branca$^{10,11}$
% ...

\vspace{4mm}

\noindent
$^{1}$ Department of Physics and Astronomy, University of South Carolina, Columbia, SC 29208, USA \\
$^{2}$ Department of Physics and Astronomy, University of California, Los Angeles, CA 90095, USA \\
$^{3}$ INFN -- Laboratori Nazionali di Legnaro, Legnaro (Padova) I-35020, Italy \\
$^{4}$ INFN -- Sezione di Bologna, Bologna I-40127, Italy \\
$^{5}$ Dipartimento di Fisica, Sapienza Universit\`{a} di Roma, Roma I-00185, Italy \\
$^{6}$ INFN -- Sezione di Roma, Roma I-00185, Italy \\
$^{7}$ Department of Physics, University of California, Berkeley, CA 94720, USA \\
$^{8}$ INFN -- Sezione di Genova, Genova I-16146, Italy \\
$^{9}$ INFN -- Sezione di Milano Bicocca, Milano I-20126, Italy \\
$^{10}$ INFN -- Sezione di Padova, Padova I-35131, Italy \\
$^{11}$ Dipartimento di Fisica e Astronomia, Universit\`{a} di Padova, I-35131 Padova, Italy \\
% ...

\begin{history}
\received{Day Month Year}
\revised{Day Month Year}
\end{history}


\begin{abstract}
\teod bolometers have been used for many years to search for neutrinoless double beta decay in \tect. CUORE, a tonne-scale  \teod detector array, recently published the most sensitive limit on the half-life, $T_{1/2}^{0\nu} > 1.5 \times 10^{25}\,$yr, which corresponds to an upper bound of $140-400$~meV on the effective Majorana mass of the neutrino. While it makes CUORE a world-leading experiment looking for neutrinoless double beta decay, it is not the only study that CUORE will contribute to in the field of nuclear and particle physics. As already done over the years with many small-scale experiments, CUORE will investigate both rare decays (such as the two-neutrino double beta decay of \tect and the hypothesized electron capture in $^{123}$Te), and rare processes (e.g., dark matter and axion interactions). 
This paper describes some of the achievements of past experiments that used \teod bolometers, and perspectives for CUORE.

\end{abstract}

\ccode{PACS numbers:}


\section{Introduction}\label{sec:intro}
A CUORE-like \teod bolometer is a particle detector that has a relatively broad dynamic range for energy detection and can achieve excellent energy resolution. In typical operating conditions, its threshold can be as low as few keV with a resolution of \ca 1~keV~FWHM, while the highest detectable energy can go up to 10-20 MeV with a resolution of \ca 5~keV~FWHM.

These devices were developed in the 90's~\cite{Alessandrello_1994,Array4_1996} and have since been used for almost 30 years to search for neutrinoless double beta (\bbz) decay in \tect, a transition that -- if observed -- would prove the Majorana nature of the neutrino and consequently the non-conservation of the lepton number. 

The high natural isotopic abundance of \tect, the intrinsic radiopurity of \teod crystals, and the energy resolution comparable to that of germanium diodes, are among the main advantages of using \teod bolometers for a \bbz decay search. Moreover, CUORE recently demonstrated the scalability of this technique with the successful operation of an array of 988 \teod bolometers installed at Laboratori Nazionali del Gran Sasso (LNGS), L'Aquila, Italy. Thanks to these qualities, \teod bolometers were and are used to explore a number of other rare processes, from two neutrinos double beta (\bbd) decay to the investigation of dark matter and axion interactions. 

In this paper we will give an overview of the studies done in the past as well as those that are on the way or that might be possible in the near future. In Sec.~\ref{sec:exp} we introduce the detector working principle with a brief description of the past experiments based on \teod. In Sec.~\ref{sec:te} we discuss results on \bbd and \bbz decays of Te isotopes, in Sec.~\ref{sec:raredec} we discuss other rare decays ($^{123}$Te, electron decay, etc.) and finally in Sec.~\ref{sec:rarepro} we discuss the possible investigation of rare processes like dark matter interaction and supernova neutrino detection.

\section{The CUORE detector}\label{sec:exp}
The word ``bolometers,'' despite being reserved for devices that measure the power of incident radiation, is used - in the jargon of particle physics - to indicate a phonon-mediated single-particle detector: a single crystal cooled to very low temperature, where an attached phonon-sensitive sensor measures the energy deposited by a particle.  If the signal is read-out after complete thermalization of the phonons, the detector is sensitive to all the energy deposited by the particle and not only to a fraction of it, as it is in the case of a more conventional ionization-based detector. This implies a number of advantages both on the achievable energy resolution and on the capability of detecting different types of particles. These devices can efficiently operate only at a very low temperature and are also called Low-Temperature Detectors (LTDs). LTDs were suggested as high-resolution soft X-ray detectors in 1984 by D. McCammon and collaborators~\cite{McCammon1984}. In the same year, E. Fiorini and T.O. Niinikoski proposed the use of bolometers for neutrino physics~\cite{Fiorini1984}. After more than 20 years of development, the reliable performance, the excellent achievable energy resolution and the scalability make these devices perfectly suitable for the use in rare event physics. 

\subsection{Low Temperature Calorimeters} 
\begin{figure}[!t]
  \begin{center}
    \includegraphics[width=0.7\textwidth]{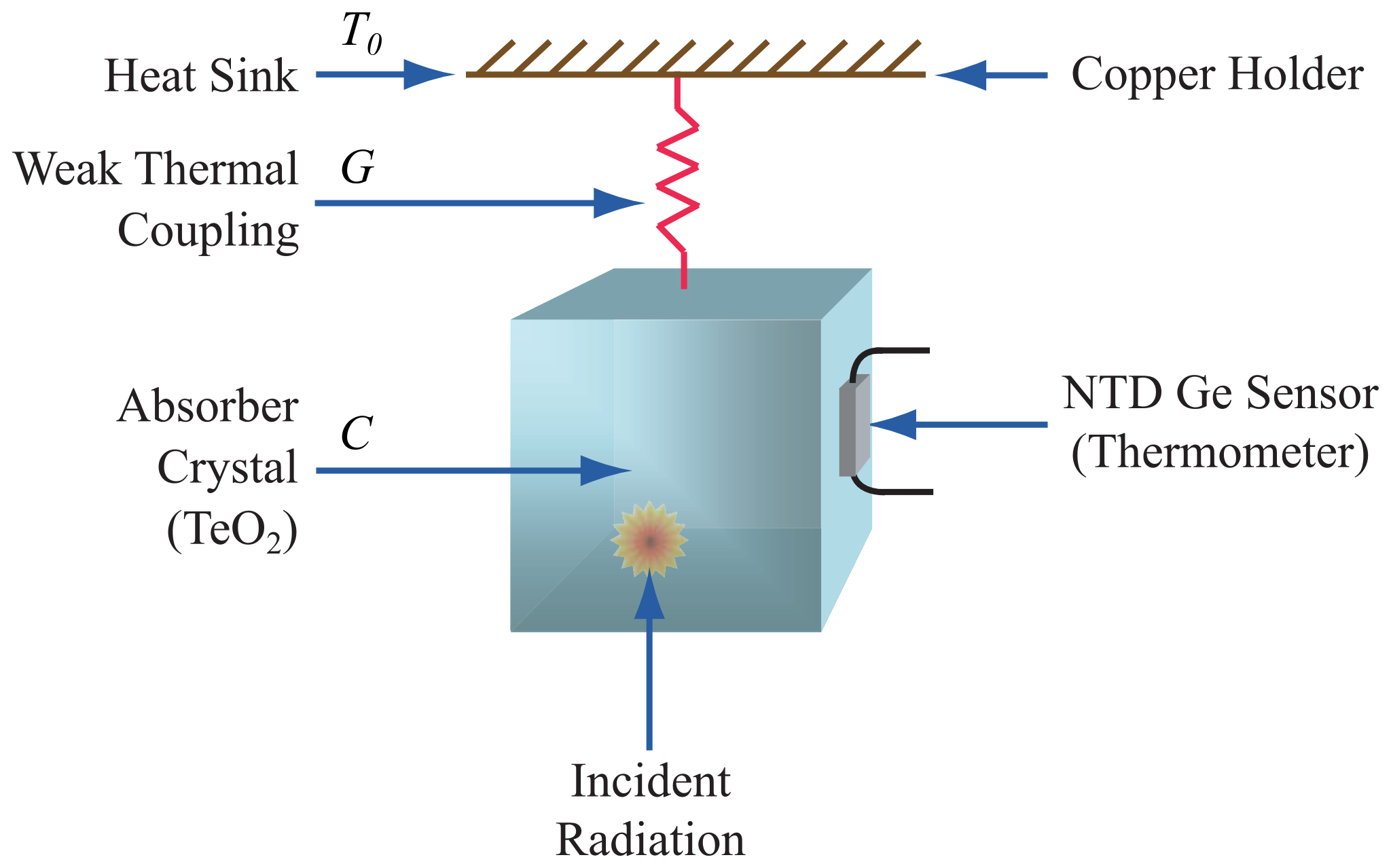} %
    \caption{Schematic of a low temperature detector. A CUORE calorimeter consists of an absorber connected to a heat sink through a weak thermal link, and read out by a temperature sensor attached to the absorber. Figure edited from Alduino {\it et al.}~\cite{Q0-detector}}
    \label{fig:ltd}
  \end{center}
\end{figure}

A bolometer consists of three main components (Figure \ref{fig:ltd}): an energy absorber, in which the energy deposited by a particle is converted into phonons, a temperature transducer (sensor) that converts phonons into a readable electric signal, and a (weak) thermal link that connects the absorber to the thermal reservoir (or heat sink) used to keep the detector at low temperature. 

In a very simplified model a device detecting thermal phonons can be represented as a calorimeter: a heat capacity $C$ connected to the heat sink (with constant temperature $T_0$) through a thermal conductance $G$. When a particle deposits an energy $E$ in the absorber, a temperature rise equal to $E/C$ is produced. Denoting the absorber temperature at time $t$ with $T(t)$ and assuming that temperature variations are small enough such that $C$ and $G$ can be considered constant, we can express the temperature evolution as :

\begin{equation}\label{eq:evolution}
  \Delta T(t)\equiv|T(t)-T_0|=\cfrac{E}{C}\,\,\mathrm{e}^{-t/\tau}\quad\mbox{with}\quad\tau\equiv\cfrac{C}{G}
\end{equation}
\noindent 
To obtain a measurable temperature rise the heat capacity $C$ of the absorber must be very small. LTDs are then operated at a temperature of about 10\,mK and are made of a superconductor or dielectric crystals, since their heat capacity is only due to the lattice contribution and follow the Debye law: $C \propto (T/\Theta)^3$ at low-temperatures.  
The decay constant, $\tau$, has to be long enough to avoid signal truncation (since in real detectors the temperature rise time is finite) but short enough to prevent pile-up. 


The intrinsic energy resolution of a conventionaò detector is primarily determined by the statistical fluctuation of the number $N$ of elementary excitations contributing to the signal. In case of a thermal-phonon detector, the fluctuations are due to the continuous phonon exchange between the absorber and the heat sink~\cite{McCammon1984}. By assuming that these fluctuations are due to a Poisson process with $\Delta N =\sqrt{N}$, the intrinsic energy resolution is proportional to the mean energy of the thermal phonons exchanged ($\varepsilon = k_\mathrm{B}T$):

\begin{equation}
\Delta E = \varepsilon\Delta N = \xi~T~\sqrt{k_\mathrm{B}C(T)}
\label{eq:intrinsicresolution}
\end{equation}

\noindent where $T$ is the detector operating temperature, $C(T)$ its heat capacity and $k_\mathrm{B}$ the Boltzmann constant. Finally, $\xi$ is a dimensionless factor that depends on the details of the detector and mainly on the phonon sensor ($\xi \geq 1$). Eq.~\ref{eq:intrinsicresolution} refers to the ``thermodynamic limit'' of the bolometer energy resolution, when all other sources of noise or energy deterioration are absent or negligible, and is independent of the incident energy itself. For \bb decay searches using TeO$_2$, we are still far from achieving this limit since the bolometer resolution is dominated by multiple``extrinsic'' noise sources.  These include the noise contributions from the cryogenic system, electronics noise from the signal read-out chain, electromagnetic interferences and microphonic noise. Eventually, this results in an energy resolution of the order of \textperthousand\ in the MeV energy range for macro-bolometers (absorber mass $\simeq 0.1 - 1\,$kg ), still an exceptionally good value when compared to ionization or scintillation based detectors that make bolometers appealing for many applications.

Finally, a further remarkable property of LTDs is that the absorber can incorporate the radioactive source being studied. This advantage is exploited in investigations of both $\beta$ decay (as in the direct measurement of the neutrino mass~\cite{MARE, HOLMES, ECHo}) and double beta ($\beta\beta$) decay.

\subsection{CUORE \teod bolometers}\label{sec:teod}

In CUORE, the energy absorber is a \teod crystal, which also acts as the source of the events of interest (\tect \bbz decays). This configuration is referred to as ``source = detector approach" and is widely used in \bbz experiments. This approach has two main advantages: a simple design of the detector (with a reduction of passive elements that might be sources of dangerous backgrounds) and a very high detection efficiency (since the decay under study occurs inside the detector). \teod crystals have a relatively high Debye temperature ($\Theta_D = (232 \pm 7)$\,K)~\cite{Barucci2001} and, hence, a small heat capacity at low temperatures. The crystals can have excellent radiopurity and are suited for ultra-low background experiments. 

\begin{figure}[!t]
  \begin{center}
    \includegraphics[width=1\textwidth]{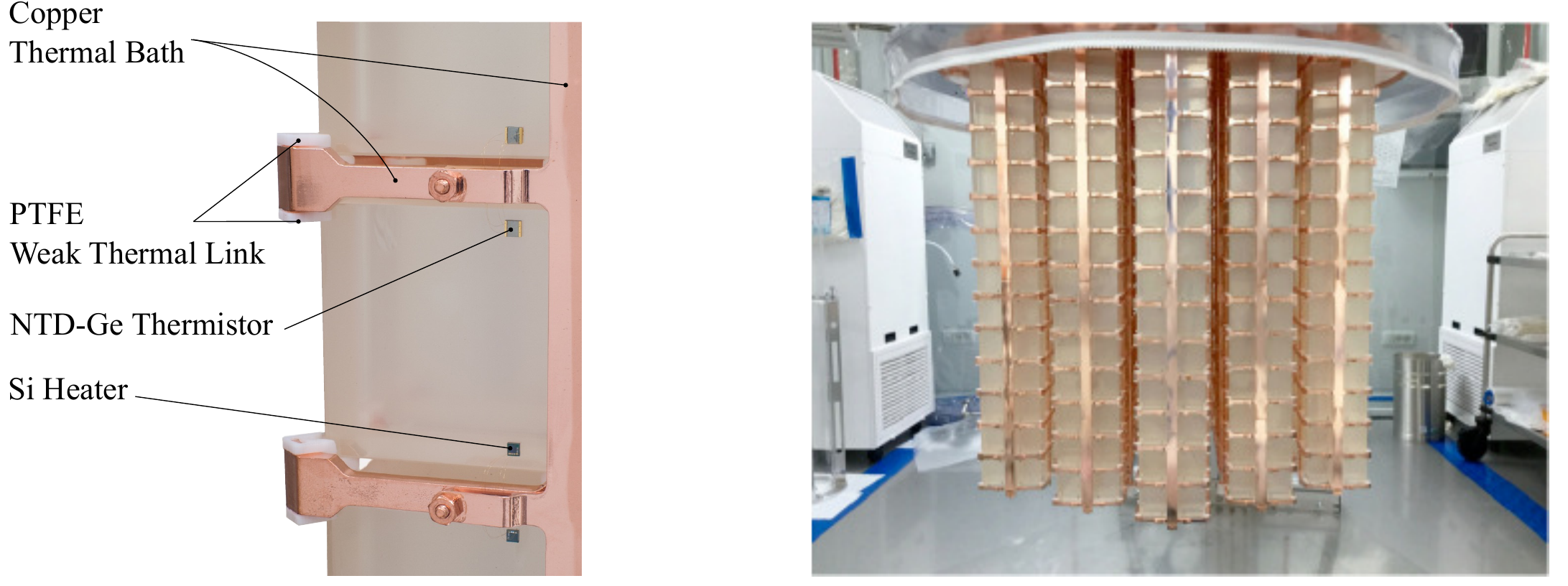} %
    \caption{(Left) Zoomed in view of a CUORE tower highlighting the fundamental components of a bolometer. (Right) The 19 towers of the CUORE detector just before the closure of the cryostat.}
    \label{fig:qdect}
  \end{center}
\end{figure}

CUORE crystals are cubic ($5 \times 5 \times 5$\,cm\textsuperscript{3}) and have an average weight of $\sim$ 750\,g (Figure \ref{fig:qdect}, left). They are made with \textsuperscript{natural}Te and each crystal contains about 208\,g of \textsuperscript{130}Te  (isotopic abundance $\eta=34.167\,\%$)~\cite{Fehr2004}. The Shanghai Institute Of Ceramics, Chinese Academy of Sciences (SICCAS) produced the crystals for CUORE. The production followed a strict crystal growing and handling protocol, specially developed for CUORE,  with the goal of  achieving a bulk contamination of $<$ $10^{-14}$\,g/g for both \textsuperscript{238}U and \textsuperscript{232}Th, and a surface contamination of $<10^{-8}$\,Bq cm$^{-2}$ for both \textsuperscript{238}U and \textsuperscript{232}Th~\cite{JCG312}. 

Each crystal is instrumented with a Neutron Transmutation Doped (NTD) germanium sensor~\cite{Wang1990,Haller94}. These devices convert the temperature change in the crystals, (typically, 0.1\,mK per MeV of deposited energy), to a resistance change according to the Shklovskii-Efros law~\cite{Efros1975}:

\begin{equation}\label{eq:selaw}
R(T)=R_0\cdot e^{\sqrt[]{(T/T_0)}}
\end{equation}

\noindent where $T_0$ depends on the doping level (related to the neutron fluence), while $R_0$ depends on both the doping level and the geometry.  The steep dependence of NTD resistance on the temperature in Equation \ref{eq:selaw} makes this material a very sensitive thermometer.  By continuously measuring the resistance of an NTD chip, it is possible to detect sudden changes in the temperature of the absorber with very high sensitivity. 

Each crystal absorber is also instrumented with a heavily doped Si chip which acts as a Joule heater (Figure \ref{fig:qdect}, left). The heaters have a nearly constant resistance at low-temperatures ($\sim$ 300\,k$\Omega$). By pulsing these resistors with a reference voltage, it is possible to impart a known amount of energy in the detector and periodically monitor its gain~\cite{98SiHeater,98SiHeater_2}.

\subsection{CUORE and its precursors~\cite{CuorePrecursors}}
The first measurements with TeO\textsubscript{2} bolometers were carried out with single detector: 6\,g and a 21\,g crystals in 1991~\cite{GIULIANI_1991}, then 34\,g in 1992~\cite{Alessandrello_1992}, 73\,g in 1993\,~\cite{Alessandrello_1993} and 334\,g in 1994~\cite{Alessandrello_1994}. The first array of TeO\textsubscript{2} calorimeters, made by four $3 \times 3 \times 6$\,cm\textsuperscript{3} crystals, each with an average weight of 340\,g~\cite{Alessandrello95} was constructed in 1994. This was the first cryogenic detector with over 1\,kg of mass.

In 1997, a tower made of 20 bolometers (5 floors of 4 crystals each) was assembled and cooled at LNGS. Each absorber was a $3 \times 3 \times 6$\,cm\textsuperscript{3} crystal and the total active mass of the array was about 6.8\,kg. The experiment, later named MiDBD (Milan Double Beta Decay), was the largest operating cryogenic detector of its time. It showcased an improvement over the previous measurements by reducing the background rate by a factor of 10, and achieving an energy resolution in the range of 5--15\,keV (FWHM) for the 2615\,keV \textsuperscript{208}Tl line\cite{Arnaboldi_2002}. The experiment placed a lower limit of $T^{1/2}_{0\nu}\geq 2.1\cdot 10^{23}$\,yr on the \bbz half-life of \textsuperscript{130}Te.

In 1998, Prof. E. Fiorini  proposed the idea of CUORE (Cryogenic Underground Observatory for Rare Events), where an array of $\sim$1000 detectors, each with a mass between 0.5 and 1\,kg, would be used to search for \bbz decay in \textsuperscript{130}Te~\cite{Fiorini_1998}. While \bbz decay remains the focus of the CUORE experiment, the large target mass and the ultra-low background would be used to search for exotic and rare decays, as well as study the interactions of WIMPS and solar axions in our detectors.  

\subsubsection{Cuoricino}
The first step towards CUORE was the Cuoricino experiment. The detector consisted of 44 large-size crystals ($5 \times 5 \times 5$\,cm\textsuperscript{3}) and 18 smaller crystals ($3 \times 3 \times 6$\,cm\textsuperscript{3}) coming from the MiDBD array. Crystals were arranged in 13 floors, with 11 modules housing 4 large-size crystals each and 2 modules housing 9 small-size ones. The total mass of TeO\textsubscript{2} was 40.7\,kg, corresponding to about 11\,kg of \textsuperscript{130}Te~\cite{qino-2008}.

Cuoricino took data from April 2003 to June 2008 for a total exposure of 19.75~kg(\tect)$\cdot$yr. It achieved an effective mean  energy resolution of ($5.8 \pm 2.1$)\,keV (FWHM) at the 2615\,keV \textsuperscript{208}Tl line. The background rates in the Region Of Interest (ROI: 2474-2580\,keV) and in the $\alpha$-region (2700-3900\,keV) were $(0.153 \pm 0.006)$\,\rate~and $(0.110 \pm 0.001)$\,\rate, respectively~\cite{qino-2008,qino-2011}. 

Cuoricino proved the feasibility of scaling to a 1000-crystals low-temperature array, achieving a good energy resolution and efficiency. Nevertheless, its performances were limited by the non-uniformity of the detector response and by the background rate in the ROI, attributed primarily to two main sources: radioactive contamination of the cryostat or its $\gamma$ shields and degraded $\alpha$ particles from the surfaces of the crystal holder (mainly the copper).  

An intense R\&D activity was pursued within the CUORE collaboration to mitigate the background sources of Cuoricino. We defined a strict protocol for crystal growth and polishing~\cite{JCG312} and adopted an elaborate cleaning and storing procedure for each part of the detector~\cite{TTT}. We designed and realized a new detector assembly line (CTAL, CUORE Tower Assembly Line ) to ensure single detector reproducibility while avoiding recontamination~\cite{CTAL}.

\subsubsection{CUORE-0}

CUORE-0~\cite{Q0-detector} was a single CUORE-like tower made of 52 natural TeO\textsubscript{2} $5 \times 5 \times 5$\,cm\textsuperscript{3} cubic crystals, arranged on 13 floors, each floor having 4 crystals. The detector had 39\,kg of active mass, corresponding to about 10.8\,kg of \textsuperscript{130}Te. The tower was constructed using the same procedures and materials selected for CUORE, it was installed in late 2012 in the former Cuoricino cryostat at LNGS and cooled to an operating temperature of T$_{0}\approx$12\,mK. The main goal of the experiment was to validate the cleaning and assembling procedures designed for CUORE and demonstrate the improvement in the background rate with respect to Cuoricino.
The detector collected data from March 2013 until March 2015. After an initial optimization phase~\cite{Q0-InitialPerformances}, the final exposure of \textsuperscript{130}Te was 9.8\,kg$\cdot$yr~\cite{Q0_PRL}.

CUORE-0 trigger thresholds ranged from 30\,keV to 120\,keV. 
The effective resolution (exposure-weighted harmonic mean energy resolution) at the 2615\,keV  \textsuperscript{208}Tl line was ($4.9 \pm 2.9$)\,keV. The recorded background rate was $0.058 \pm 0.004\mbox{(stat)}\pm 0.002\mbox{(syst)}$\,\rate~ in the ROI, with an improvement of about factor 3 with respect to Cuoricino, and $(0.016\pm 0.001)$\,\rate~ in the $\alpha$-region, roughly 7 times smaller than in Cuoricino. Both the improvements are mainly ascribed to a reduction in the amount of degraded $\alpha$ particles coming from the holder, with the result that the background rate in the ROI of CUORE-0 was dominated by the irreducible contamination in \textsuperscript{208}Tl of the cryostat.

The new assembly line and procedures resulted in robust and reproducible detector characteristics, and the detector showed excellent bolometric performances with an effective background suppression. These results demonstrated that all the solutions and procedures selected for CUORE-0 could meet the goals for CUORE.

\subsubsection{CUORE}

CUORE~\cite{Q_Proposal,Q_NIMA,Q_AHEP} consists of a closely packed array of 988 \tect crystals (5 $\times$ 5 $\times$ 5 cm$^3$ each) organized in 19 towers, each identical to the CUORE-0 one (i.e. 13 floors with 4 crystals each). It is, by far, the largest detector operated as a low-temperature calorimeter.  The array is hosted in one of the largest cryostat ever constructed to reach a base temperature of $<$\,10\, mK~\cite{Q_Cryostat_4K,Q_Cryostat_10mK}. 
Even from the point of view of radioactivity the cryostat is one of the significant differences between CUORE-0 and CUORE. One of the major sources of $\gamma$-ray background in CUORE-0 was indeed the cryostat itself. By carefully selecting the materials for the cryostat construction, we hoped to have mitigated the $\gamma$-ray background substantially for CUORE.

The full detector assembly took almost two years, from September 2012 to July 2014. The successful completion of the task definitively proved the effectiveness of the CTAL, gluing and bonding protocols. Before being installed inside the CUORE cryostat, the towers were stored inside the CUORE clean room into sealed containers constantly flushed with clean N\textsubscript{2} gas to prevent any contamination from radon, waiting for the end of the commissioning of the cryogenic system. The tower installation was successfully performed in summer 2016, and CUORE started its commissioning phase at the beginning of 2017. The initial months were dedicated to the system optimization and setting of the optimal working points.

For the initial phase of CUORE, we operated the detectors at 15\,mK, where we found the signal-to-noise ratio from the thermistors to be optimal. After a total TeO\textsubscript{2} exposure of 86.3\,kg$\cdot$yr, the preliminary background rate in the ROI is $(0.014 \pm 0.002)$\,\rate~, with a dramatic improvement with respect to CUORE-0. Concerning detector performances, in this first phase an effective mean energy resolution, on 984 of 988 functioning channels, of $(7.7\pm 0.5)$\,keV has been obtained. Additional optimization campaigns are in progress with the aim to improve the detector performances by optimizing the experimental operating conditions. The first CUORE preliminary science results have been recently published~\cite{Q_PRL}.

\subsection{Detector read-out and acquisition}
We present here a brief description of the read-out chain and the data acquisition systems adopted in both CUORE-0 and CUORE (the systems used in the previous experiments were quite similar as similar was data production, later described).
To read out the signal, each thermistor is biased in differential configuration with a bipolar voltage generator connected to a pair of load resistors, finally connected to the thermistor terminals. The resistance of the thermistor varies in time with the temperature, $R(t)$, and the voltage across it, $V_R(t)$, is the bolometer signal. The value of the load resistors, $R_L$, is chosen to be much higher than $R(t)$ so that $V_R(t)$ results only dependent to $R(t)$, following the law introduced in Eq. \ref{eq:selaw}. The signal $V_R(t)$ is amplified by a low-noise front-end electronics~\cite{Q_front_end}, filtered by a 6-pole active Bessel-Thomson filter~\cite{Q_bessel} and then digitized with an 18-bit analog-to-digital converter~\cite{GIACHERO_PHD,DIDOMIZIO_PHD} (ADC) with a sampling rate of 1\,kHz (125\,Hz in CUORE-0). The front-end boards, which provide the bias voltage, the load resistors, the amplifier, and the filter boards are placed outside the cryostat, at room temperature.



\subsection{Data production}
A software derivative trigger is used to identify thermal event pulses and collect them in 5\,s windows. Each window is divided into two parts: 1\,s before the trigger and 4\,s after it. The period before the trigger (pre-trigger) is used to establish the baseline temperature of the crystal, and the remaining 4\,s is used to determine the pulse amplitude. Forced random triggers are used to evaluate the detector noise and to perform threshold studies. As mentioned in Sec. \ref{sec:teod},  a Joule Si-heater resistor is also coupled to each crystal with epoxy and is used to generate reference thermal pulses
every 300\,s. These induced pulses are used to stabilize the gain of the bolometer against temperature fluctuations. To improve the signal-to-noise ratio, an optimal filter is applied to each pulse~\cite{Gatti1986}, exploiting the distinct power spectra of particle-induced and noise waveforms. The pulse amplitude is determined from the maximum value of the filtered waveform. The data are grouped into datasets, which last approximately one month. At the beginning and at the end of each dataset, the detector is exposed to a radioactive source to calibrate the detector. The CUORE calibration system consists of 12 low-intensity \textsuperscript{232}Th sources attached to Kevlar strings~\cite{Q_DCS}. During the calibration, the strings are lowered from room temperature into the cryogenic volume amongst and around the towers. In CUORE-0 and its precursors, calibration procedure were similar but the source was external to the cryostat.

The data production is a procedure that converts the data from a series of triggered waveforms into a calibrated energy spectrum. The procedure developed for CUORE is similar to the one developed for CUORE-0~\cite{Q0-analysis}, but scaled to 1000 channels. First, a time-coincidence analysis is performed to search for events due to external sources that deposit energy across multiple bolometers.  Then, a series of event selection cuts are performed to maximize the sensitivity to physics events. This selection includes the identification of cryostat instability and malfunction periods and a series of pulse shape cuts to reject deformed or non-physical events. The searched-for events are usually confined within one crystal. However, many background sources deposit energy in multiple crystals within the response time of the detector. By event multiplicity, it is possible to form multiplets of events that occur within a coincidence window of $\pm 5$\,ms in different crystals. For each multiplet it is possible to build the related spectrum:
\begin{itemize}
\item\mspec spectrum is the energy spectrum of all events, each hit crystal corresponds to an entry in the spectrum and no coincidence criteria are applied;
\item\mspecone spectrum is the energy spectrum of the events with the requirement that only one bolometer is involved (multiplicity 1 or \mspecone events);
\item\mspectwo spectrum is the energy spectrum of the events with the requirement that two bolometers triggered (multiplicity 2 or \mspectwo events);
\item\summspectwo spectrum is the energy spectrum associated to \mspectwo  pairs, each pair produces an entry with an energy $E(\Sigma_2)$ that is the sum of the energies of the two events.
\end{itemize}
Higher-order multiplets are used only to evaluate the contribution to the background from muons. The signal cut efficiency as a function of energy is defined as the fraction of true signal events that pass all the event cuts. More details on these selection techniques are presented in dedicated papers~\cite{Q0-analysis,Q0_2nu}. 

To optimize the search for low energy rare events~\cite{Q_low}, such as solar axions or WIMP scattering, the standard derivative trigger is replaced by an optimum trigger (OT), that allows us to lower the threshold below 30\,keV. The OT algorithm used in CUORE is an improved version to that presented in the past~\cite{OT}. The data buffer is divided into slices that are continuously filtered in the frequency domain with the optimal filter mentioned above. The filtered waveforms have an improved signal-to-noise ratio, and baseline fluctuations are reduced. This allows triggering on the filtered signal in the time domain with a threshold as low as $< 10$\,keV. The optimum-triggered waveforms are converted into physics data following a procedure similar to the standard data production introduced above. 

\subsection{Dangerous backgrounds and mitigation strategies}
The search for a rare event, such as a \bbz decay source, means operating a detector avoiding, as far as possible, any spurious sources that can mimic the event under study (e.g. radioactive sources or cosmic rays). Any event producing an energy deposition similar to the searched-for decay increases the background and hence spoils the sensitivity of the experiment. Background sources are cosmic rays, natural or artificial radioactive contaminations in the laboratory environment and in the experimental set-up, and finally the radio-impurities contained in the detector itself. A combination of approaches is necessary to reach a low background rate. First an underground location is fundamental to get rid of cosmic rays. Shielding the detector components from environmental radioactivity, screening materials to achieve low radioactive contamination, carefully preventing recontamination with radon daughters, and actively rejecting background events are all also required.

CUORE, as its precursors, is operating in Hall A of the LNGS laboratory. This provides the detector with an overburden of 1400\,m of rock (3600\,meters of water equivalent) which reduces the cosmic ray rate by 6 orders of magnitude relative to the surface. The measured residual rate is roughly one muon per hour in CUORE~\cite{CuoricinoMuons}. The CUORE detector is completely surrounded (laterally, below and above) by lead shields. In particular, the innermost one is made of ancient Roman lead. This is characterized by an extremely low radioactivity and it is almost fully depleted of \textsuperscript{210}Pb. The outer shields protect the detector from environmental $\gamma$-rays and neutrons. The innermost shields protect the detector from the $\gamma$ radiation produced in the outer shields or in the cryostat itself and have to meet strict radioactivity limits in order to be a negligible source of background. 

The contaminants present in the detector raw materials were reduced through a careful screening. Three main techniques for screening materials were used: direct $\gamma$-ray and $\alpha$-particle spectroscopy, inductively coupled plasma mass spectroscopy (ICP-MS), and neutron activation analysis (NAA). The CUORE towers were built using clean and controlled techniques. All activities were carried out in a dedicated class 1000 (ISO 6) cleanroom located in the underground CUORE hut. The cleanroom contained glovebox-enclosed systems for assembling the towers in radioclean conditions under nitrogen atmosphere. The assembly procedure used two separate workstations: one for gluing chips to crystals and one for building and instrumenting the towers. At the end, the tower installation was performed in a controlled clean room environment with radon-free air, filtered by a dedicated system~\cite{RADON_FREE}.


\section{Second order weak decays of Te}\label{sec:te}
 \bbd decay is a second-order weak decay, which occurs in a few of the even-even nuclei, where the initial nucleus (A, Z) can decay into its isobar (A, Z+2), emitting two electrons and two anti-neutrinos in the process. For a fraction of these nuclei, the decay to neighboring odd-odd isobar (A, Z+1) is either energetically disfavored or forbidden by the spin-parity conservation lawslaws, and the experimental observation is easier. The process conserves lepton number and is consistent with the ``Standard Model" of particle physics. \bbd decay has been observed for 11 nuclei~\cite{Barabash}. However, if the final states have only two electrons and no anti-neutrinos are emitted, then we can have another rare mode of decay which is known as \bbz decay which violates lepton number by 2 units ($\Delta L$=2). Both the decay modes can occur either to the $0^+$ ground state or to the excited states ($0^+_i, 2^+, 2^+_i$) of the final nuclei. Similarly, a second-order weak transition can happen from (A,Z) to (A, Z-2) through \bpbp, \bpEC or \ECEC decay modes. 
In \bpbp transitions a significant fraction of the available energy goes to the positron masses. The transition energies as well as the experimental sensitivities are correspondingly lower. 

Naturally occurring tellurium contains three candidate isotopes which can undergo second order weak decay: $^{130}$Te (\bmbm), $^{128}$Te (\bmbm) and $^{120}$Te (\bpbp/\bpEC/\ECEC). $^{130}Te$ and $^{120}Te$ have relatively high $Q$ values and, hence, have a higher expected decay rate. The sensitivity of the search is further enhanced by the fact that there are less background sources to contend with at higher energies. In the following subsections, we will highlight the results for the (\bb) decay searches from CUORE-0 and CUORE.

\subsection{\tect neutrinoless decay}\label{sec:130te0nu}
The search for \bbz decay is mainly motivated by its ability to probe the Majorana nature of the neutrinos. The existence of \bbz decay necessarily implies a Majorana mass for the neutrino, regardless of the mechanism that is used to induce \bbz decay in a gauge theory~\cite{blackbox_theorem, blackbox_theorem_2}. The importance of Majorana neutrino mass stems from the fact that it can not only help explain the smallness of the observed neutrino masses~\cite{seesaw_Minkowski,seesaw_Mohapatra}, but it also provides a mechanism to explain the matter-antimatter asymmetry in the universe~\cite{leptogenesis_Fukugita,leptogenesis_Bari}. Further, the search for \bbz decay also probes the absolute mass of the neutrinos and their hierarchy~\cite{RMP_Frank}. The decay rate of \bbz, assuming light Majorana neutrino exchange, takes the form 
\begin{equation}\label{eq:decay-rate}
  (T^{1/2}_{0\nu})^{-1} = {\nmfn} |\langle m_{ee} \rangle|^2
\end{equation}
where $G^{0\nu}$ is the phase-space factor the nuclei, $M^{0\nu}$ is the nuclear transition matrix element, and $\langle m_{ee} \rangle = |\Sigma_i U_{ei}^2 m_i| $ is the effective Majorana mass for light neutrinos and is the only  relevant parameter measured in \bbz decay. $m_i$ denotes the neutrino mass and $U_{ei}$ is the corresponding component of the neutrino mixing matrix. The sensitivity of the search depends on how precisely we can measure $T^{1/2}_{0\nu}$.

The crystal size of CUORE ensures that both the electrons emitted in the \bbz decay of \tect are mostly contained inside a single detector crystal, unless the event occurs near the surface. In fact, Monte Carlo simulations place the containment efficiency to be $\sim$88\%. Therefore, the experimental signature that we are looking for is a mono-energetic peak at the Q-value of the \tect decay, $Q_{\beta\beta}$ = 2527.515~$\pm$~0.013~keV~\cite{Q_value_Redshaw,Q_value_Scielzo,Q_value_Rahaman}.  In presence of background-fluctuations, the sensitivity of the experiment can be expressed as 
\begin{equation}\label{eq:sensitivity}
  T^{1/2}_{0\nu} \propto \eta \cdot \epsilon \cdot \sqrt[]{\frac{M\cdot t}{B \cdot \delta E}}
\end{equation}
where $\eta$ is the isotopic abundance of the candidate nuclei, $\epsilon$ is the efficiency of the detector, M is total active mass of the detector, $t$ is the livetime of the experiment, $B$ is the background rate expressed in \rate, and $\delta E$ is the energy resolution in the ROI. As mentioned before, bolometers have excellent energy resolution and are suited for a sensitive \bbz search. On the other hand, considerable of effort is needed to reduce the background rate in the ROI. While the \Qbb~lies in between the 2615-keV line from $^{208}$Tl and its Compton edge, the primary background contribution to our ROI is from degraded $\alpha$ particles from materials near the surface of the detector and the multi-Compton scattered $\gamma$-rays, namely, from $^{208}$Tl and $^{214}$Bi. 

\subsubsection{CUORE-0} 

CUORE-0 was designed to validate the background rejection techniques that were developed for the construction of the CUORE detector. It was the first of the twenty towers to have come out of the CUORE detector assembly line. Despite being a prototype for CUORE, CUORE-0 has been a sensitive experiment on itself, searching for \bbz in \tect with a total exposure of 35.2 kg$\cdot$yr of TeO$_2$, or 9.8 kg$\cdot$yr of \tect. 
In the following subsections, we will first discuss the results from CUORE-0 before dwelling on the results from CUORE, pointing out the subtle differences between the two experiments wherever necessary. 

\begin{figure}
\begin{center}
\includegraphics[width=0.9\textwidth]{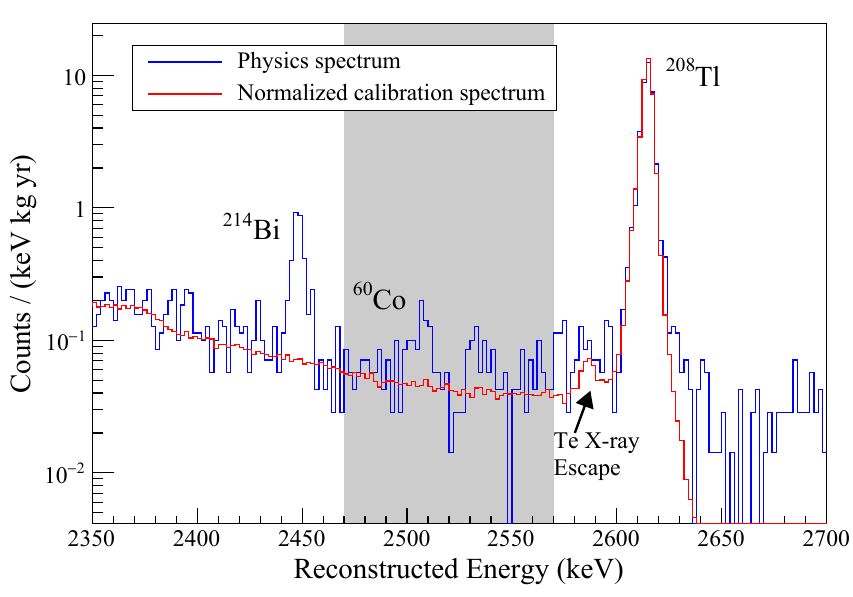} %
\caption{The CUORE-0 spectrum around the region of interest. The calibration data (red) has been normalized to the physics data (blue) at 2615 keV to elucidate the shape of the background in the ROI. The shaded region corresponds to the energy range used in the ROI fit.}
\label{fig:Q0_NormalizedSpectrum_ROI}
\end{center}
\end{figure}

\begin{figure}[h]
\begin{center}
\includegraphics[width=0.9\textwidth]{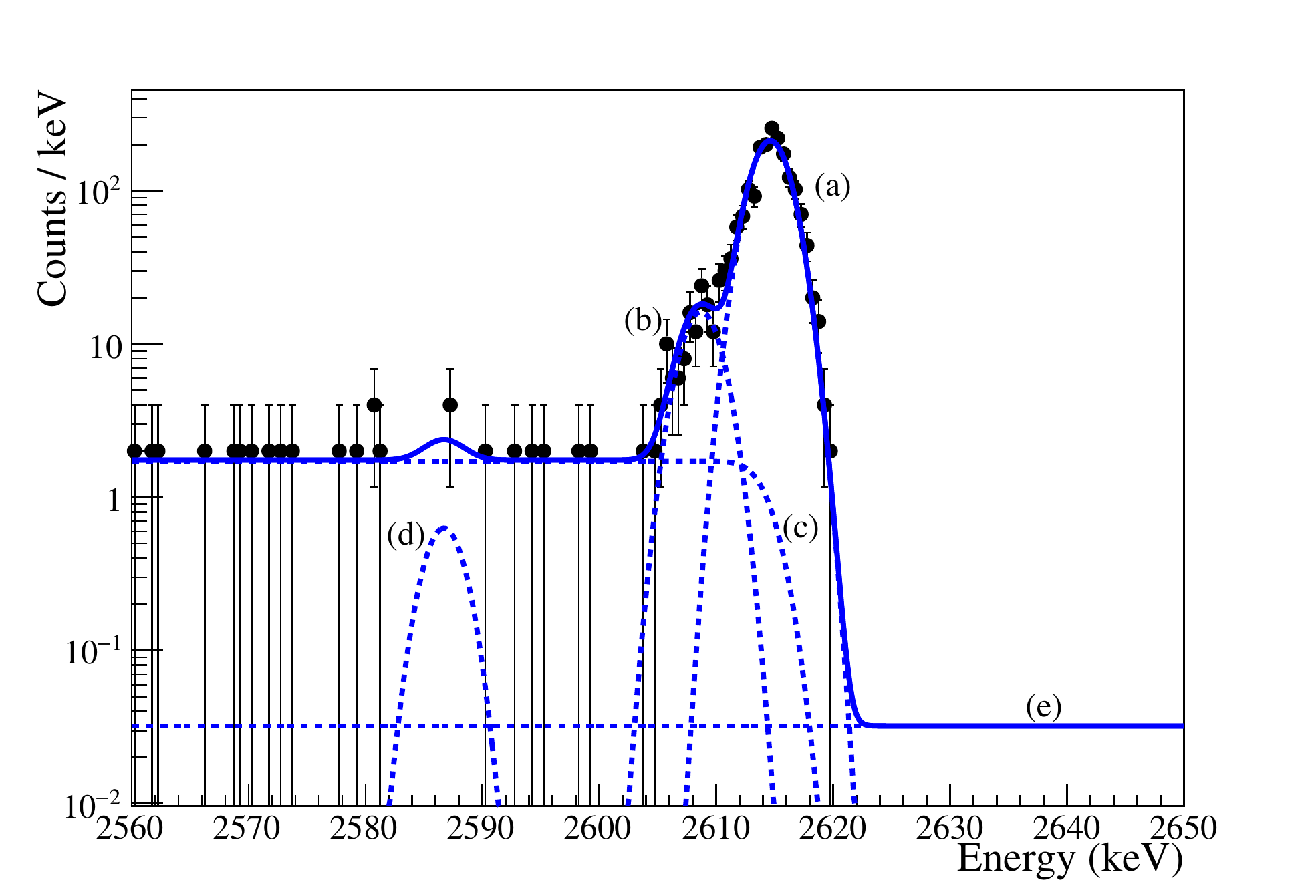} %
\caption{A 5-component fit to the 2615\,keV calibration peak for a single detector in a CUORE-0 dataset: (a) A primary Gaussian for the main peak; (b) a secondary Gaussian at a slightly lower energy; (c) the smeared step-function; (d) the x-ray escape peak; and (e) a flat background. The primary and secondary Gaussians are considered to be the detector's response to a mono-energetic peak and are defined as the lineshape of the detector.}
\label{fig:Q0_SingleChannelResponse}
\end{center}
\end{figure}

The analysis techniques and results from CUORE-0 have been well described in the literature~\cite{Q0-analysis}. We will simply cite the results here. Figure~\ref{fig:Q0_NormalizedSpectrum_ROI} shows the CUORE-0 spectrum around the ROI which we defined as the energy range 2470$-$2570 keV. One can clearly see that the shape of the  background in the ROI closely matches the shape of the normalized calibration data. This indicates that the background in the ROI is dominated by the multiple scattered $\gamma$ background rather than a flat $\alpha$ background as expected from degraded $\alpha$'s. The ROI also shows a peak at $\sim$ 2507 keV which is attributed to the sum peak of $^{60}$Co gammas. The presence of $^{60}$Co is due to the activation of the copper frames and internal shielding in the cryostat. The $^{60}$Co sum line is still far away from the Q-value of \bbz, but needs to be taken into account while fitting the ROI. 
To fit a peak in the region of interest it is necessary to understand the response of each detector to a mono-energetic \bbz decay. We modeled the response based on the measured response to the 2615 keV calibration peak. The full calibration response was modeled using a five component un-binned extended maximum likelihood (UEML) fit to 2615 keV line for each detector in a dataset (see Figure~\ref{fig:Q0_SingleChannelResponse}). 
The mono-energetic peak is well described by a double-Gaussian. The secondary Gaussian has a slightly lower mean energy and accounts for $\sim$5\% of the events under the peak. The reason for the presence of a secondary peak is still under investigation. The substructure was independent of the data analysis steps and the double-Gaussian reproduced the data well across the energy range.  

\begin{figure}[h]
\begin{center}
\includegraphics[width=1\textwidth]{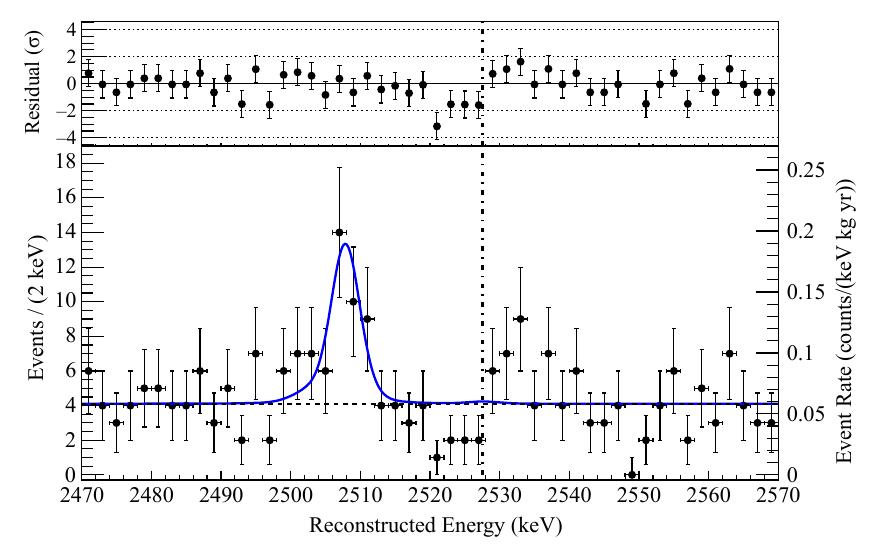} %
\caption{The simultaneous UEML best-fit to the data ($\mathcal{M}_1$) in the region of interest for CUORE-0. The dashed-dotted line shows the position of the \Qbb. The total \tect exposure was 35.2\,kg$\cdot$yr. Figure from K. Alfonso \textit{et al.}~\cite{Q0_PRL}.}
\label{fig:Q0_ROI_Unblinded}
\end{center}
\end{figure}

Once the lineshape is known, we fit the ROI to determine the yield of \bbz events. We performed a simultaneous UEML fit in the ROI (Figure~\ref{fig:Q0_ROI_Unblinded}). The fit components included a double-Gaussian at the Q-value, another double-Gaussian accounting for the $^{60}$Co sum peak, and a constant background attributed to multi-Compton events from $^{208}$Tl and $\alpha$ decays at the surface.

We obtained a best-fit decay rate of~\cite{Q0_PRL} 
\begin{equation}\label{eq:Q0-best-rate}
  {\hat\Gamma_{0\nu}} = [0.01 \pm 0.12\text{(stat)} \pm 0.01\text{(syst)}] \times 10^{-24} \text{yr}^{-1}, 
\end{equation}
and a background rate in the ROI of 
\begin{equation}\label{eq:Q0-bkg}
  \mathrm{b_{ROI}} = [0.058 \pm 0.004\text{(stat)} \pm 0.002\text{(syst)}] ~ \text{counts/(keV$\cdot$kg$\cdot$yr)}. 
\end{equation}

The 90\%\,C.L. median sensitivity for half-life, assuming no \bbz signal and the above background, was calculated to be 2.9$~\times$~10$^{24}\,$yr. We obtained a Bayesian upper limit on the decay rate by using a profile likelihood method.
We got an upper limit of $\hat\Gamma_{0\nu}< 0.25\times 10^{−24}\,$yr$^{-1}$ or $T^{1/2}_{0\nu}> 2.7 \times 10^{-24}\,$yr at 90\%\,C.L. The probability to obtain a more stringent limit is 55\% and the limit is slightly worse than the median sensitivity because of the slight upward fluctuation in the background at the Q-value.

\subsubsection{CUORE}

\begin{figure}
\begin{center}
\includegraphics[width=0.9\textwidth]{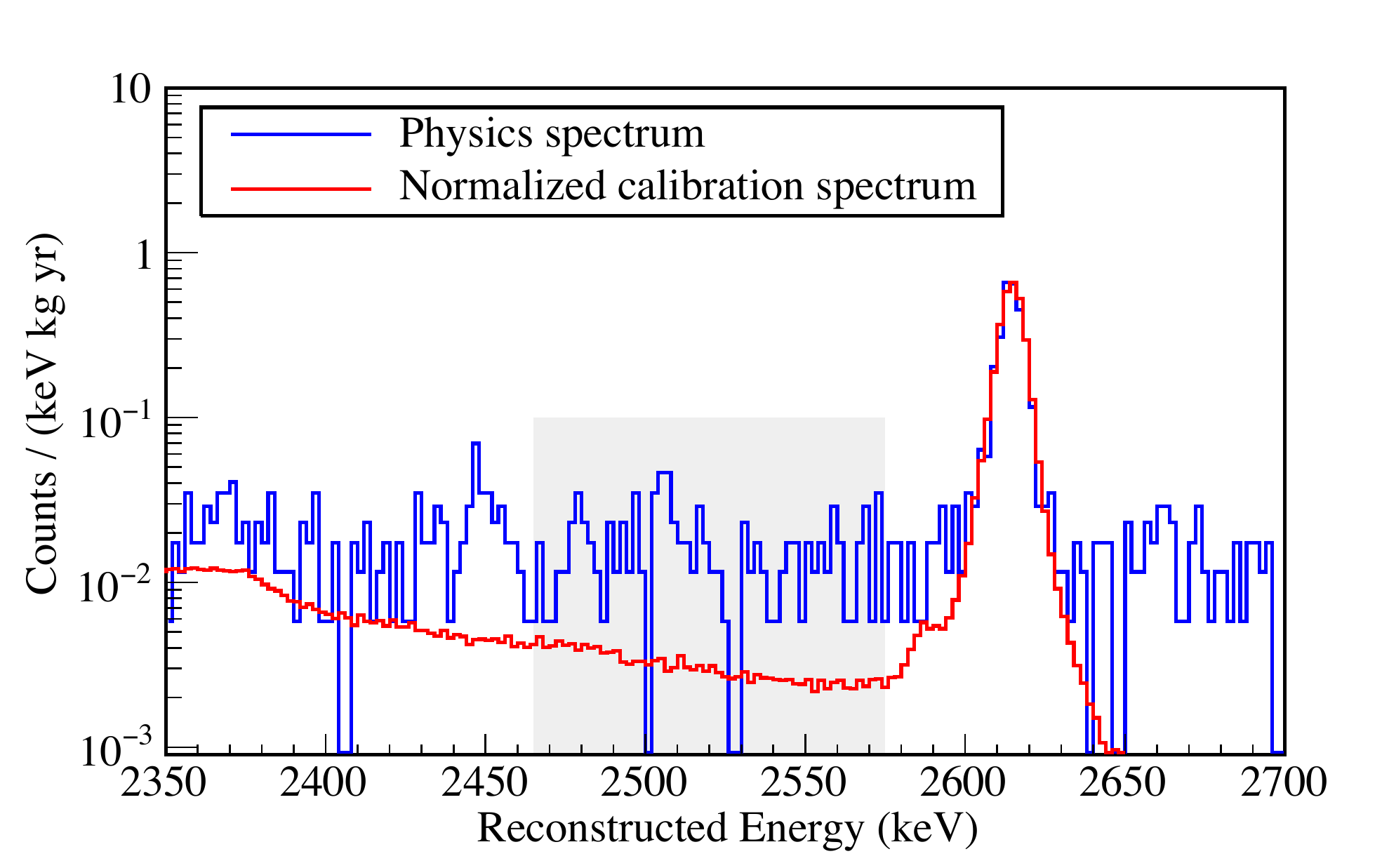} %
\caption{The CUORE spectrum around the region of interest (exposure 86.3 kg$\cdot$yr). The calibration data (red) has been normalized to the physics data (blue) at 2615 keV to elucidate the shape of the background in the ROI. The shaded region corresponds to the energy range used in the ROI fit.}
\label{fig:Q_NormalizedSpectrum_ROI}
\end{center}
\end{figure}

The physics spectrum for CUORE~\cite{Q_PRL}, close to the region of interest, is shown in Figure~\ref{fig:Q_NormalizedSpectrum_ROI}. We see that the background in the ROI is relatively flat and does not follow the shape of the Compton-scattered background as expected from $\gamma$-rays. This is an important distinction to make from CUORE-0; we did manage to significantly reduce the $\gamma$-ray background, but are limited by the background from the degraded $\alpha$s near the detector surfaces. 

\begin{figure}
\begin{center}
\includegraphics[width=0.9\textwidth]{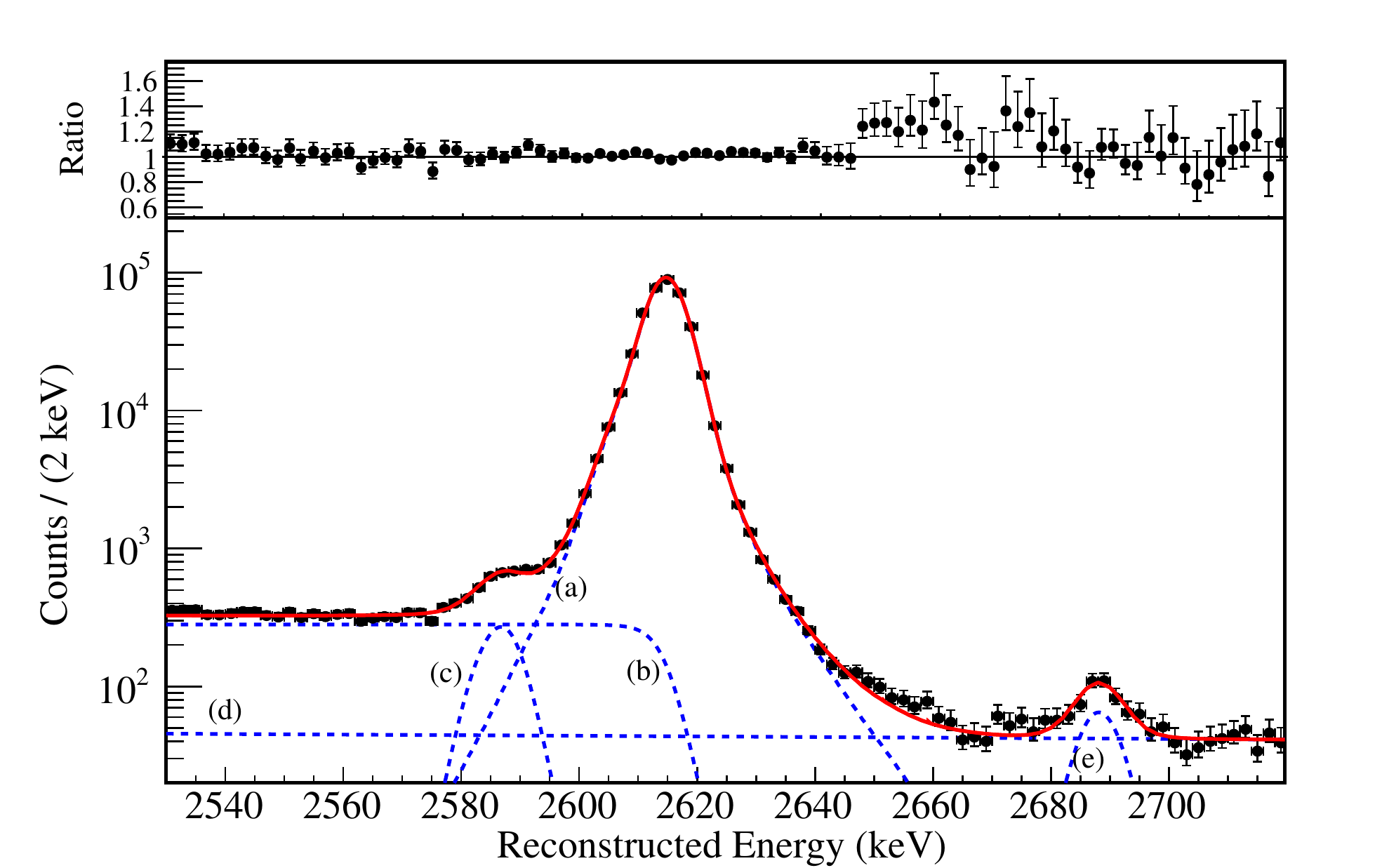} %
\caption{2615 keV lineshape of all the 19 towers of CUORE summed together. The red line is the sum of the fits of all individual channels while the blue-dashed lines shows the different components of the lineshapes. The subcomponents were identified as (a) multi-Gaussian peak describing the main photopeak, (b) contribution from multi-scatter Compton events, (c) a peak ascribed to the escape of 27-31~keV X-rays from Te after a 2615~keV deposition, (d) flat background due to degraded $\alpha$s, and (e) a peak attributed to the single escape peak of the coincident absorption of 2615~keV and 583~keV $\gamma$-rays. Figure from C. Alduino \textit{et al.}~\cite{Q_PRL}.}
\label{fig:Q_Response}
\end{center}
\end{figure}

We have also found the response of the CUORE detectors to be different from that of CUORE-0. Unlike CUORE-0, where a double-Gaussian was used to fit a photopeak, in CUORE the same peak was better described by a triple-Gaussian fit, with subpeaks on the left and right of the main peak respectively. We are currently investigating the origin of these substructures. However, we did take the lineshape uncertainty into account and treated it as systematic uncertainty in our final calculations. Figure~\ref{fig:Q_Response} shows the combined fit for the 2615\,keV photopeak of all the channel-dataset pairs in CUORE. 

\begin{figure}
\begin{center}
\includegraphics[width=0.9\textwidth]{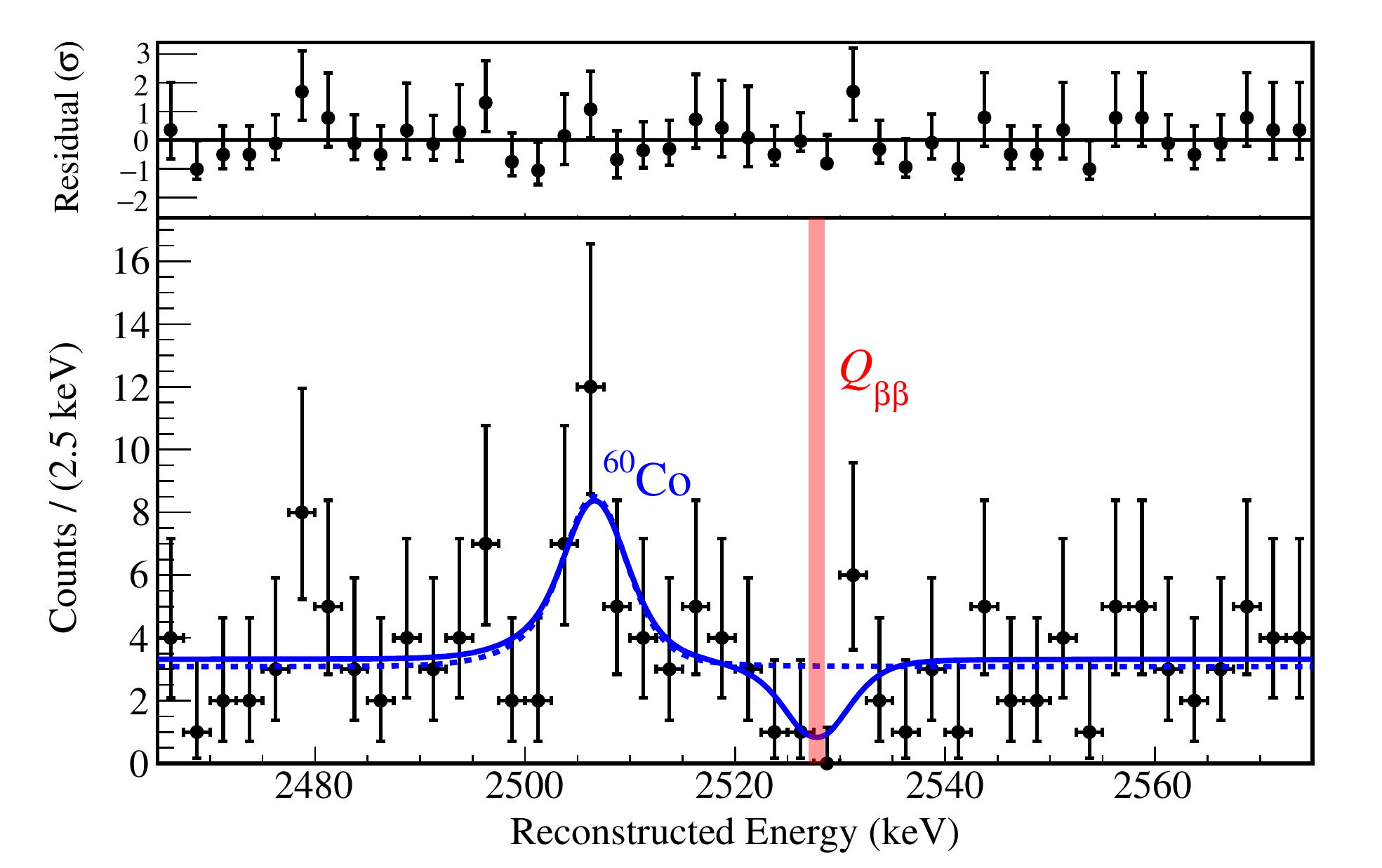} %
\caption{The simultaneous UEML best-fit to the data ($\mathcal{M}_1$) in the region of interest for CUORE. The systematic uncertainty of the reconstructed energy at \Qbb is represented by the width of the vertical band. A total \tect exposure of 86.3\,kg$\cdot$yr was used for the analysis. See text for details. Figure from C. Alduino \textit{et al.}~\cite{Q_PRL}.}
\label{fig:Q_ROI_Unblinded}
\end{center}
\end{figure}

Following the procedure outlined for CUORE-0, we performed a simultaneous UEML fit in the ROI (2465$-$2475 keV) to determine the yield of \bbz events (Figure~\ref{fig:Q_ROI_Unblinded}). The fit components included a triple-Gaussian at the Q-value, another triple-Gaussian accounting for the $^{60}$Co sum peak, and a continuum background attributed to multi-Compton events from $^{208}$Tl and $\alpha$ decays at the surface. Figure~\ref{fig:Q_ROI_Unblinded} shows the 155 events that passed all the selection criteria together with the UEML fit. We observed a large negative fluctuation for a signal at the Q-value, leading to a best-fit decay rate of 
\begin{equation}\label{eq:Q-best-rate}
  \hat{\Gamma}_{0\nu} = [-1.0^{+0.4}_{-0.3} \text{(stat)} \pm 0.1\text{(syst)}] \times 10^{-25} \text{yr}^{-1}.  
\end{equation}
The best-fit background index in the ROI was 
\begin{equation}\label{eq:Q-bkg}
  \mathrm{b_{ROI}} = [0.014 \pm 0.002] ~ \text{counts/(keV$\cdot$kg$\cdot$yr)}. 
\end{equation}

We integrated the profile likelihood for positive values of ${\hat\Gamma}_{0\nu}$ and obtained a 90\%\,C.L. upper limit to the half-life of $T^{1/2}_{0\nu}> 1.4 \times 10^{25}\,$yr. The probability of obtaining a more stringent limit was 2\% and the median 90\%\,lower sensitivity of $T^{1/2}_{0\nu}$ was $7.0 \times 10^{24}\,$yr. A frequentist treatment of the data yielded $T^{1/2}_{0\nu}> 2.1 \times 10^{25}\,$yr at a 90\% C.L, with a lower limit sensitivity of $7.6 \times 10^{24}\,$yr (90\%\,C.L.).

Combining the negative log-likelihood curve of CUORE with that of CUORE-0 and Cuoricino, we obtained a decay-rate of ${\hat\Gamma}_{0\nu} < 0.47\times 10^{-25}\, \text{yr}^{-1}$, corresponding to a half-life of $T^{1/2}_{0\nu}> 1.5 \times 10^{25}$\,yr. A frequentist treatment of the data yielded $T^{1/2}_{0\nu}> 2.2 \times 10^{25}$\,yr (90\%\,C.L.). 

\begin{figure}
\begin{center}
\includegraphics[width=0.9\textwidth]{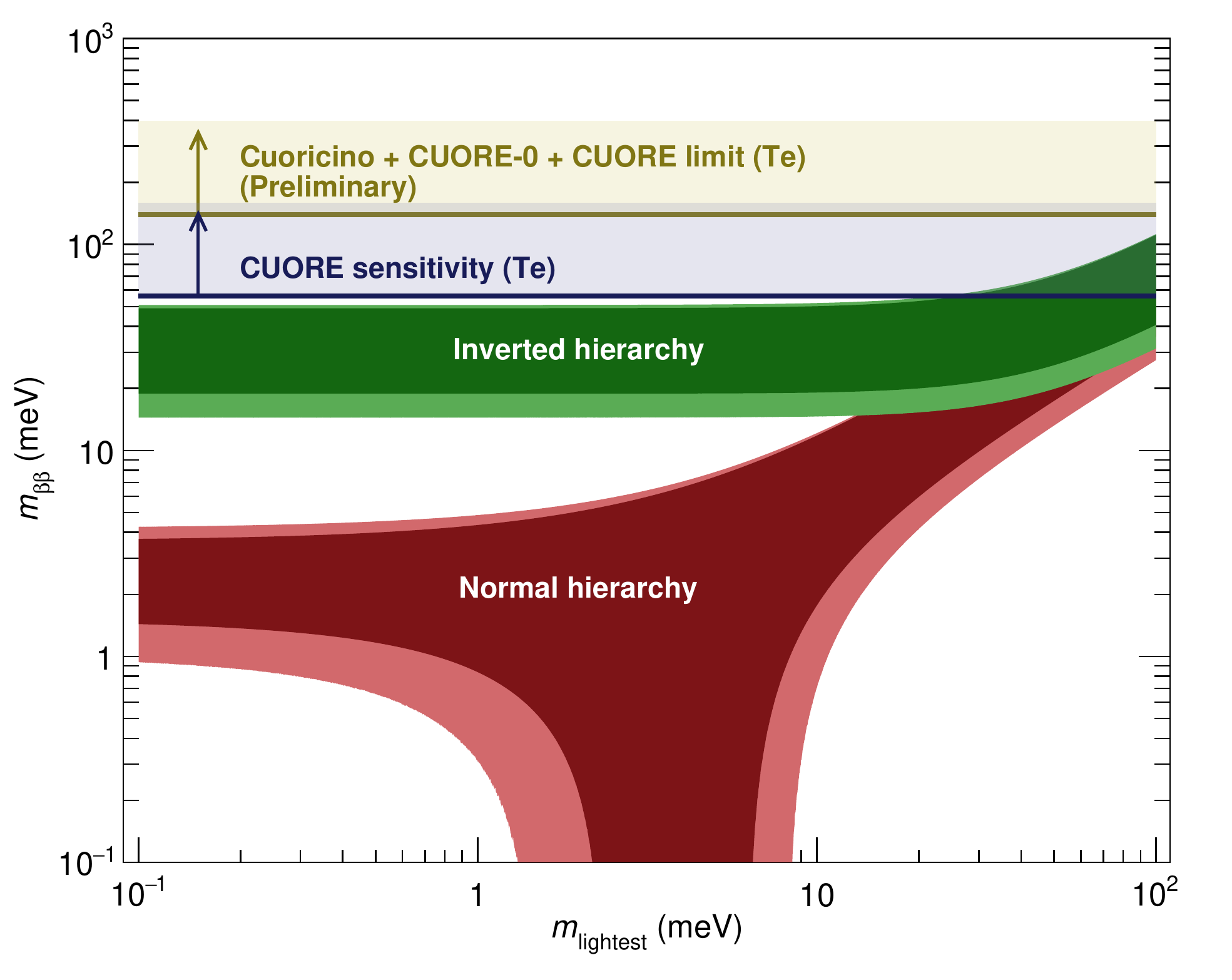} %
\caption{Dependence of $m_{\beta\beta}$ on the absolute mass of the lightest neutrino. CUORE yields an $m_{\beta\beta}$ value of $<$ (140--400) meV at 90\% C.L., depending on the nuclear matrix element. It will get very close to probing the inverted hierarchy region in 5 years of livetime.}
\label{fig:Q_M_bb_vs_lightest}
\end{center}
\end{figure}

The obtained half-life limit was used to constrain the effective Majorana mass of neutrinos in the framework of $0\nu\beta\beta$ induced by light Majorana-neutrino exchange. While the phase space factor for the decay has been precisely calculated, the nuclear matrix element calculations are model dependent. Depending on the model considered, we obtained an upper limit of $m_{\beta\beta}<$(140$-$400)\,meV (see Figure~\ref{fig:Q_M_bb_vs_lightest}). 

Efforts are going on to improve our experimental conditions and analysis techniques, through which we hope to meet our background level expectation of $\sim$0.01~counts/(kg$\cdot$keV$\cdot$yr) and an energy resolution of $<$5\,keV. This will yield a sensitivity of $T^{1/2}_{0\nu}> 9 \times 10^{25}$~yr (90\%\,C.L.) with 5 years of live time, and will be very close to probing the inverted hierarchy mass region as shown in Figure~\ref{fig:Q_M_bb_vs_lightest}. 

\subsection{\tect two neutrino decay}\label{sec:130te2nu}
In \bbd, two electron anti-neutrinos are emitted in addition to the two electrons. The decay occurs whether or not neutrinos are their own antiparticles and it is the only decay mode allowed by the Standard Model for \bb candidates like \tect. This process is the slowest nuclear decay directly observed~\cite{Exo,Kz}. It is studied because it can provide relevant insight into nuclear theory~\cite{2nCoraggio} and may as well represent an important irreducible source of background for the search of \bbz decay. Indeed, the energy resolution of the detector produces a smearing of the spectrum that, near the end point, leads to a mixing of \bbd and \bbz signals so that the former can be a relevant background source for the latter. However,\tect has a long half-life and this effect is completely negligible for bolometers~\cite{IHE} while it is sizable in poor energy resolution detectors like liquid scintillators~\cite{SNO2n}.

Bolometers are certainly among the most sensitive detectors that can be used to search for long lived decays. However, the measurement of \bbd half-life is extremely challenging, unlike the transitions with a strong signature (e.g. the monochromatic lines of $^{209}$Bi $\alpha$ decay \cite{Marcillac,NOIBiEcc}) where the measurement is quite straightforward. 
The two electrons emitted in the \bbz decay of \tect travel at most 3 mm in \teod crystals, so that in most cases both are detected in the same bolometer where the decay occurs (\mspecone signals). The resulting energy spectrum of the sum kinetic energies of the two electrons has a bell-like shape extending from 0 up to the \tect Q-value, with a maximum at about 850~keV. The signature of the decay is therefore quite weak. Moreover, a number of background sources contribute to the detector counting rate in the same region.



\subsubsection{\bbd decay in CUORE-0}
To disentangle the \bbd signal in CUORE-0 data we had to identify and estimate all the major background sources, resulting in a ``background model'' which was able to explain the counting rate of the array over a wide energy range~\cite{Q_low}. In the following subsections we will briefly describe the procedure adopted for the model construction before discussing the main results.
\begin{enumerate}
\item We developed a detailed Monte Carlo simulation of the detector, the cryogenic apparatus and the shield system, taking special care in reproducing not only the geometry of the set-up but also the response of the detector (energy resolution, threshold, etc) and the daq/analysis procedures (pile-up treatments, definition of coincidences, etc) so that simulated data closely resembled the real data;
\item We validated the simulation using a calibrated source, proving that the reproduction of experimental data is extremely good above 300~keV;
\item We defined the list of all the sources that could contribute to the counting rate of an underground experiment, from the environmental neutron, muon and $\gamma$ fluxes to the usually present U, Th and K radioactive contamination of materials. The list was completed using information from the material radioactive assay campaign preceding CUORE-0 detector construction. Complementary information also came from the analysis of the CUORE-0 spectrum itself: from the analysis of $\alpha$ peaks we reconstructed the violation of secular equilibrium in U and Th chains; from $\alpha$ and $\gamma$ peaks we identified some radioactive contaminant specific of our experiment (e.g. $^{125}$Sb likely due to Te activation, $^{190}$Pt due to crystal contamination during the growth procedure, etc);
\item We produced a set of simulations, one for each source and each possible location of the source (here we made the assumption of an uniform distribution of contaminants in bulk and/or surface of the various geometrical elements) and fitted their sum to CUORE-0 experimental data, with activities being the free parameters to be estimated by the fit;
\item We fitted experimental data with a Bayesian algorithm adopting, for source activities, \emph{priors} defined using constraints from material assays (already mentioned at point 3). The experimental data used in this analysis were \mspecone, \mspectwo and \summspectwo spectra averaged over the whole array. These three spectra were simultaneously fit to their simulated counterpart deriving \emph{posterior} estimates for the background source activities together with estimates of the correlation among the different sources;
\item We investigated possible bias in source activity estimate by observing how the fit was dependent on binning, threshold, data-sample, etc.
\end{enumerate}

\begin{figure}
\begin{center}
\includegraphics[width=1\textwidth]{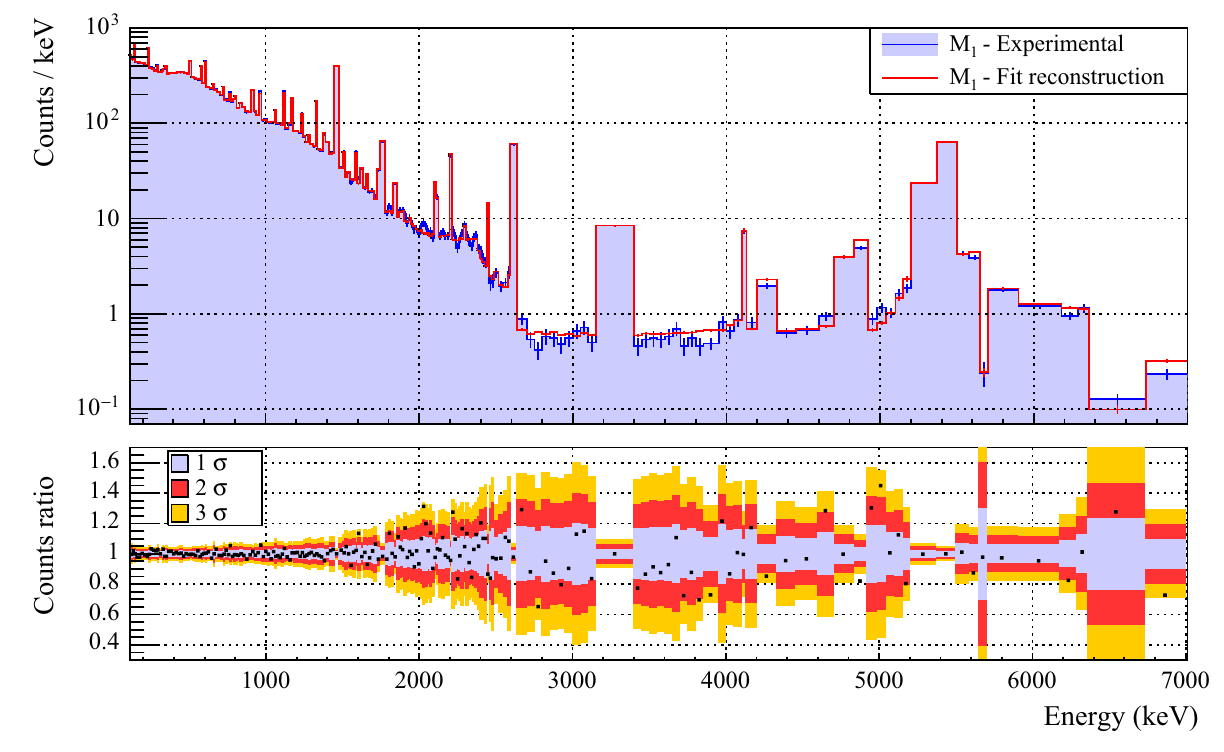} %
\caption{Comparison between the experimental \mspecone and its reconstruction (top panel). In the bottom panel the bin by bin ratios between counts in the experimental spectrum over counts in the reconstructed one are shown; the corresponding uncertainties at 1, 2, 3 $\sigma$ are shown as colored bands centered at 1. Fit residuals distribution is approximately gaussian 
with $\mu= (-0.03 \pm 0.09)$ and $\sigma=(1.1 \pm 0.1)$. Figure from C. Alduino \textit{et al.}~\cite{Q0_2nu}.}
\label{fig:JAGS-M1}
\end{center}
\end{figure}

Figure \ref{fig:JAGS-M1} shows the reconstruction of the \mspecone spectrum (\mspectwo and \summspectwo reconstructions are shown in \cite{Q0_2nu}). 
Figure \ref{fig:JAGS-2n} shows the \bbd contribution predicted by the fit. The obtained \bbd half-life is shown in Table~\ref{tab:2nu} together with those reported by previous experiments: NEMO and MiDBD.
The large systematic error gives a clear indication of the difficulties that are faced in this kind of analysis, difficulties that are all related to the identification and quantification of the competing background sources. 

\subsubsection{\bbd decay in MiDBD}
MiDBD was a small array of \teod bolometers, among which two crystals were enriched in \tect and two in \tecv (therefore being depleted in \tect). These four crystals were used to measure the \bbd half-life of \tect studying the difference between the spectra recorded by two \tect enriched and two depleted ones~\cite{Arnaboldi_2002}. Assuming a similar background recorded by the four crystals and due to contaminants isotropically distributed around them, the only difference in their counting rate could be ascribed to \tect \bbd decay in the enriched ones. 

An excess of the counting rate of the \tect enriched detectors with respect to the depleted was indeed observed. However, it was immediately evident that the condition of identical contaminants isotropically distributed in the crystals and around them was not met. This required a careful analysis of the contribution of sources violating this hypotheses, that resulted in a large systematic uncertainty (see Table~\ref{tab:2nu}).

\begin{figure}
\begin{center}
\includegraphics[width=1\textwidth]{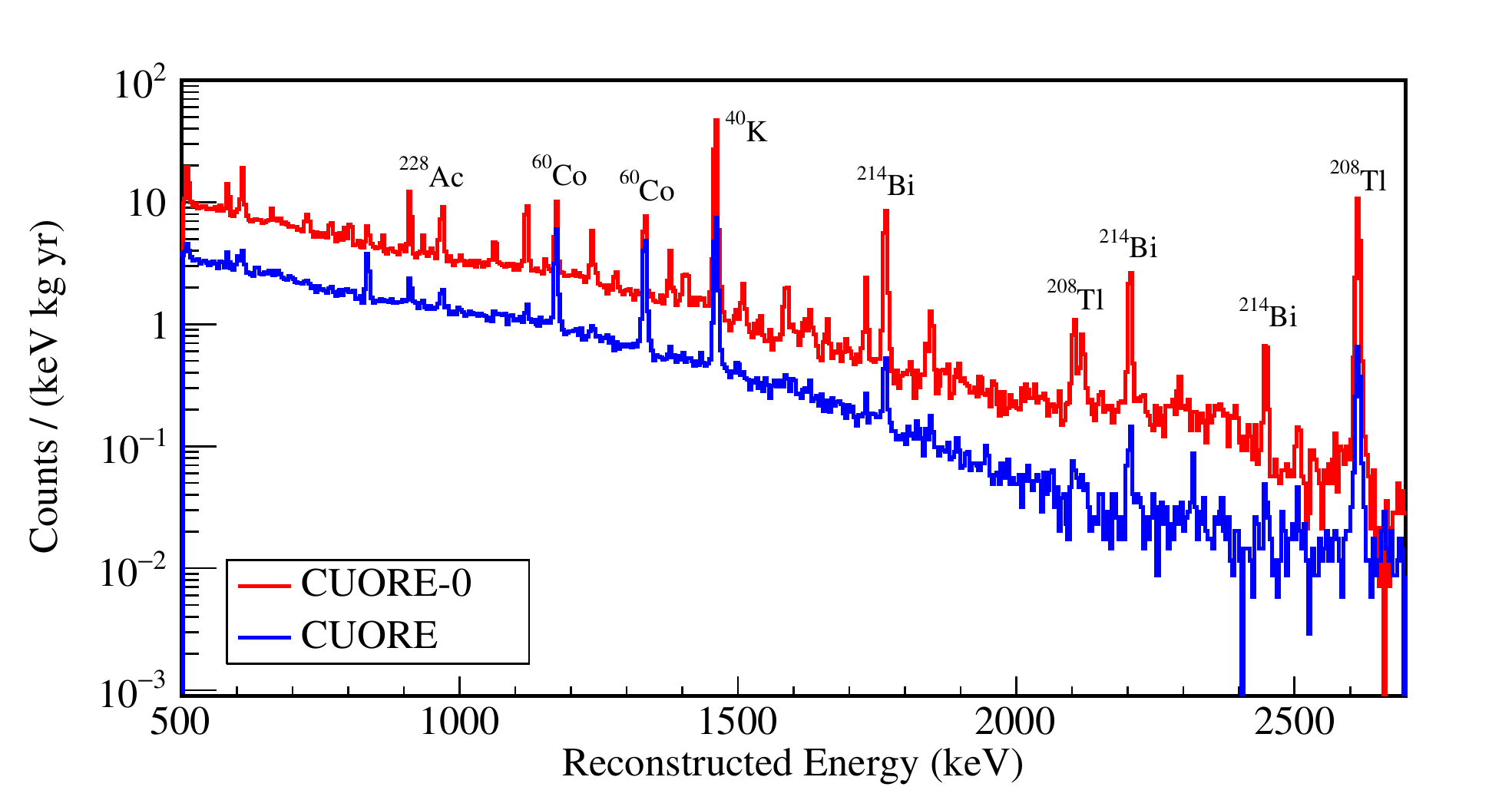} %
\end{center}
\caption{Energy spectrum of \mspecone events recorded in a 35.2 kg$\cdot $yr exposure by the CUORE-0 array and in a 86.3 kg$\cdot$yr exposure by the CUORE one. At a first sight it is clear that \bbd is not the dominant contribution to the counting rate.}
\label{fig:cuore0+cuore}
\end{figure}

\subsubsection{\bbd decay, from CUORE-0 to CUORE}

The CUORE-0 result was obtained using only natural crystals and using a full background reconstruction. This means that instead of having few free parameters as in MiDBD (the activities of the few sources that affected the counting rate of the two couple of enriched and depleted detectors differently), the whole background had to be studied and activities estimated for 57 sources. 
However, thanks to the good energy resolution and threshold of the detectors (both much improved with respect to the MiDBD ones), a larger coincidence efficiency together with an enhanced ability in reproducing detector behavior through Monte Carlo simulation, the systematic uncertainty is 5 times lower than in MiDBD. 
The main contribution to the systematic uncertainty in CUORE-0 analysis, is still due to the imperfect reproduction of the contaminant distribution topology, in and around the detectors, a weakness that very likely CUORE, thanks to the high granularity of its detector will overcome.

\begin{table}[h]
\ttbl{23pc}{Estimates of the \tect half-life obtained by CUORE-0 and MiDBD (bolometric experiments using \teod detectors) and NEMO.}
{\begin{tabular}{l l} \\[6pt]
\hline\hline
Experiment	&	Half-live [$\times$ 10$^{20}$~y]			\\
\hline
CUORE-0~\cite{Q0_2nu}	&8.2 $\pm$ 0.2 (stat.) $\pm$ 0.6  (syst.)	 \\
MiDBD~\cite{Arnaboldi_2002}	&6.1 $\pm$ 1.4 (stat.) $^{+2.9}_{-3.5}$ (syst.)	 \\
NEMO ~\cite{NEMO2n}		&7.0 $\pm$ 0.9 (stat.) $\pm$ 1.1 (syst.)\\
\hline
\end{tabular}} 
\label{tab:2nu} 
\end{table}

CUORE analysis will adopt a procedure similar to the one used for CUORE-0. Given that the mass is nearly 20 times larger and the background counting rate reduced by about a factor 4  (see Fig.~\ref{fig:cuore0+cuore}), the statistical error will accordingly be reduced. Moreover, the self shielding of the detectors implies that the inner core of the array should be less sensitive to external sources (from shielding and cryostat) with two advantages: a lower counting rate and a reduced number of sources contributing to systematics. This study is presently in progress.


\begin{figure}
\begin{center}
\includegraphics[width=1\textwidth]{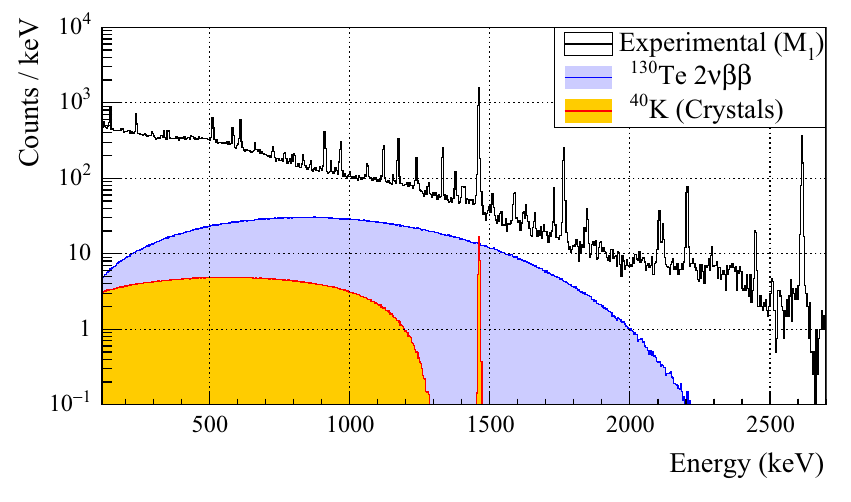} %
\caption{CUORE-0 \mspecone compared to the \bbd contribution and the radioactive source that has the strongest correlation with \bbd, \kq in the crystal bulk. Both the \bbd and the \kq contribution are normalised to the activities resulting from the reconstruction. Figure from C. Alduino \textit{et al.}~\cite{Q0_2nu}.}
\label{fig:JAGS-2n}
\end{center}
\end{figure}


\subsection{\tect decays to excited states}\label{sec:130texx}
\begin{figure}
\begin{center}
\includegraphics[width=1\textwidth]{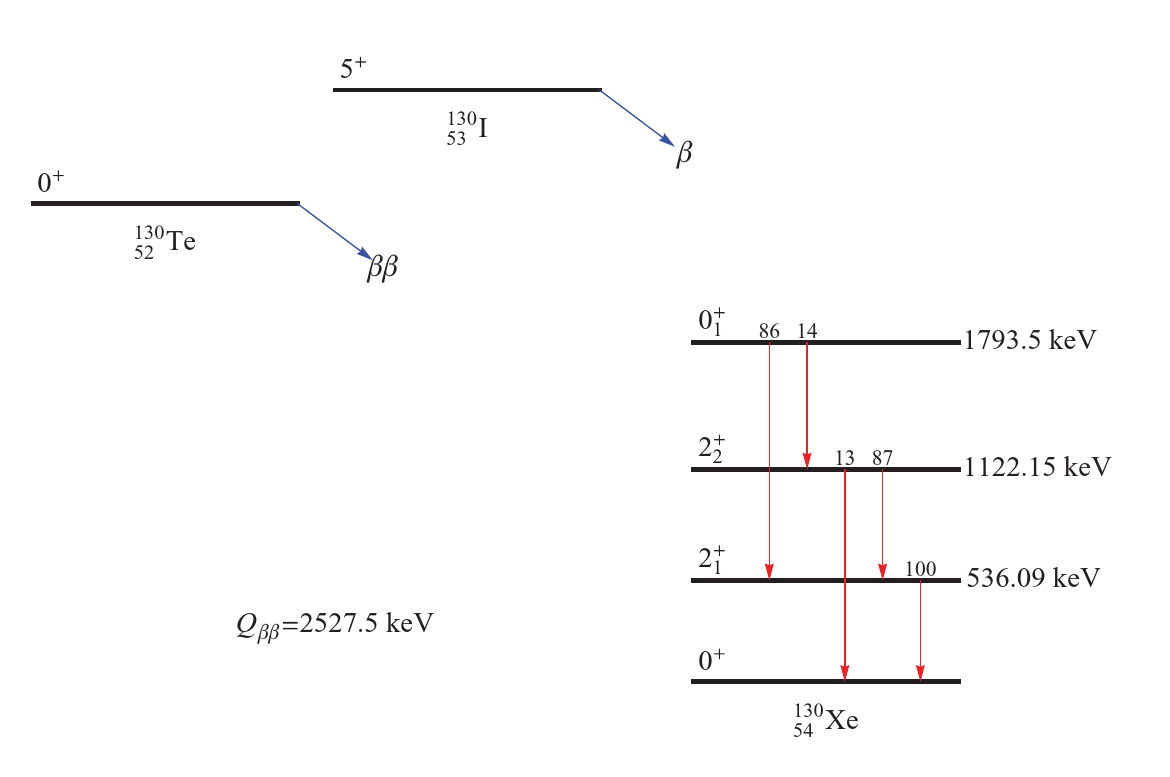} %
\caption{Decay scheme for \tect, showing the
energy levels and the branching ratios for the $\gamma$ rays}
\label{fig:TeLevels}
\end{center}
\end{figure}

\tect double beta decay may occur -- in both $2\nu$ and $0\nu$ modes -- not only as a ground-state to ground-state transition ($0^+\rightarrow 0^+$) but also as a transition from \tect ground-state to the $^{130}$Xe $0^+_1$ excited state at 1793.5~keV ($0^+\rightarrow 0^+_1$). 
This level de-excites through 3 different cascades crossing the 1122.15 keV ($2^+_2$) and the 536.09 keV  ($2^+_1$) levels as show in Figure~\ref{fig:TeLevels}. Despite the long life-time theoretically predicted for this decay channel, the interest in its study arises from the fact that in an array of close packed detectors the strong signature provided by the simultaneous detection of one or two gammas in the detectors that are close to the one where the decay occurs, can lead to a background-free search.
Less appealing is on the other hand the study of the transition to one of the two $2^+$ excited levels ($0^+\rightarrow 2^+_{1,2}$) since its amplitude is strongly suppressed by angular momentum conservation.

The strategy adopted for this analysis is to exploit multiple coincidence patterns that are combined together to extract the partial decay widths: $\Gamma^{0^+_1}_{0\nu}$ and $\Gamma^{0^+_1}_{2\nu}$.
The events considered are \mspectwo and \mspecthree, corresponding to the simultaneous energy deposition in two or three nearby bolometers. One of the bolometers is the \emph{source} of the decay. The other one or two are the \emph{spectators}: they just detect the photons escaping from the \emph{source}. The \emph{spectators} are required to record a monochromatic signal, corresponding to the full conversion of one or more $\gamma$s (this implies a dramatic reduction of the background counting rate). 

The 90\% C.L. limits obtained with Cuoricino~\cite{qino-exc} are:
\begin{equation}\label{eq:best-rate}
T_{1/2} (\beta\beta2\nu-{0^+_1}) > 1.3 \times 10^{23}\, yr
\end{equation}
\begin{equation}\label{eq:best-rate}
T_{1/2} (\beta\beta0\nu-{0^+_1}) > 9.4 \times 10^{23}\, yr
\end{equation}
A similar analysis, using CUORE-0 data and exploiting the reduced background levels and the improved performances of the detectors, has recently allowed to improve these limits ~\cite{Q0excited}.  


\subsection{$^{120}$Te decays}\label{sec:120te}
While \bmbm decay remains the most promising mode for the possible detection of neutrinoless mode in double beta decay experiments, there has been a renewed interest in other lepton number violating processes, namely, the neutrinoless modes of double positron decay (\bpbp), positron emitting electron capture (\bpEC) and double electron capture (\ECEC). These processes are interesting because they are dominated by right-handed weak currents, and can provide crucial inputs for understanding the underlying mechanism of \bbz, if observed. 
Tellurium has one candidate isotope, \tecw, for which \bpEC and \ECEC decays are allowed. The \bpbp decay in \tecw is energetically forbidden, since the mass difference between the neutral mother and daughter atoms (\tecw and \tinw, respectively) is less than 4$m_e c^2$.  On the other hand, the \ECEC decay is typically suppressed since the final state requires emission of additional particles (namely, two inner-bremsstrahlung photons) to conserve energy-momentum. That leaves \bpEC as the most favorable decay channel that can be probed experimentally with \tecw. The $Q$-value of the reaction, for \bpEC, is given by 
\begin{equation}
\begin{split}
\mathrm{Q_{EC\beta^+}} & = M\mathrm{(^{120}_{52}{Te})} - M\mathrm{(^{120}_{50}{Sn})} - 2~m_{e} c^{2} \\ 
& = \Delta M - 2~m_{e} c^{2},
\end{split}
\end{equation}
where the atomic mass difference, $\Delta M$, is 1714.8 $\pm$ 1.3~keV. We should also take into account the binding energy of the captured electron which lowers the available kinetic energy ($\mathrm{K_{EC\beta^+}}$) of the emitted positron
\begin{equation}
\begin{split}
\mathrm{K_{EC\beta^+}} & = \Delta M - 2~m_{e} c^{2} - E_{b} \\
			  & = (692.8 \pm 1.3)~keV - E_{b},
\end{split}
\end{equation}
where $E_b$ is the binding energy of the absorbed atomic electron. X-ray and/or Auger electrons are emitted in the de-excitation of the ionized daughter atom. However, the characteristic energy of these are of $O$(30~keV), and they mostly deposit their energy inside the crystal where the decay has occurred, unless the decay is near the surface. 

\begin{figure}
\begin{center}
\includegraphics[width=0.8\textwidth]{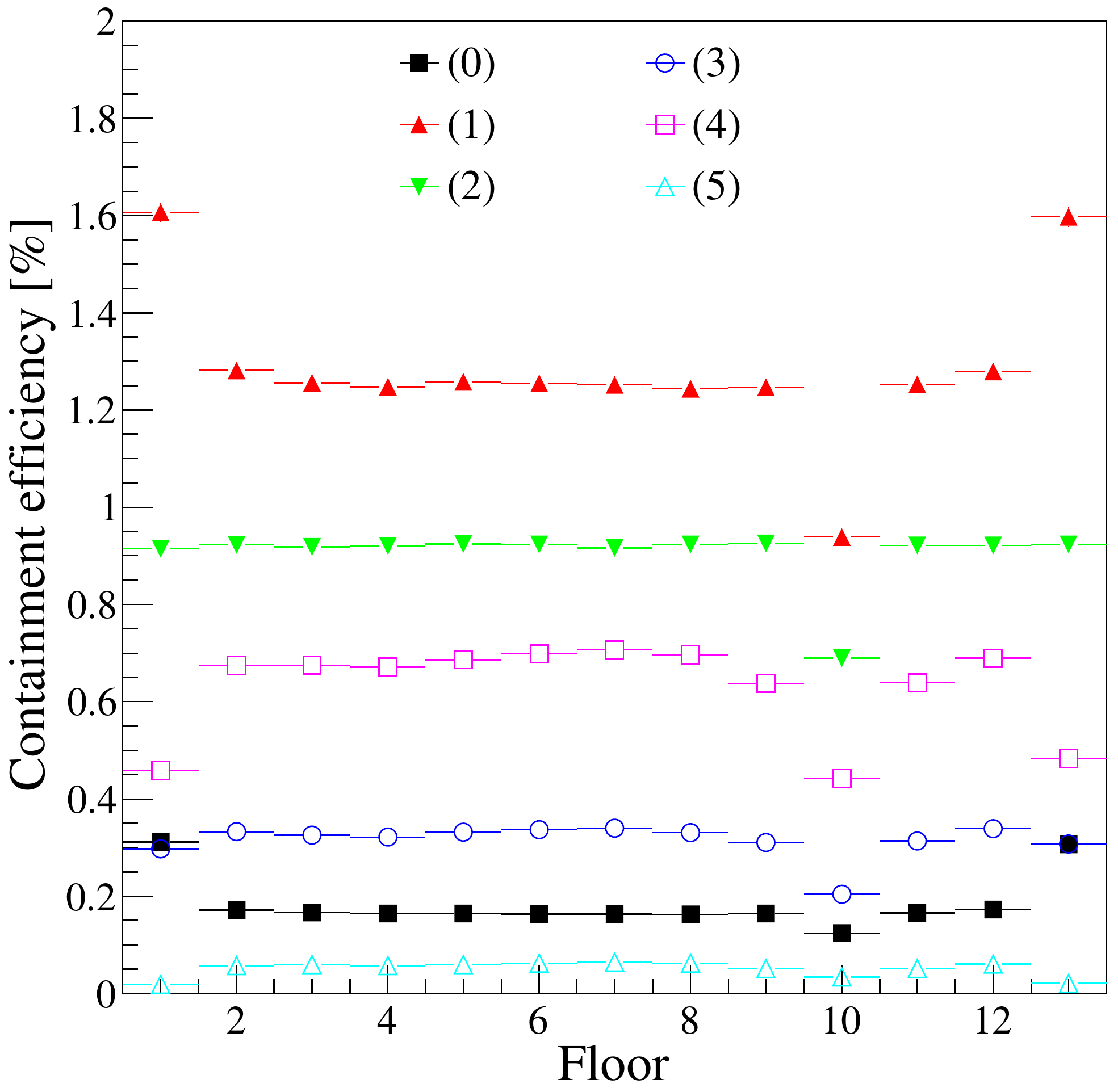} %
\caption[caption]{Detection efficiency for aforementioned signatures (see Table~\ref{tab:120te_eff}) in different floors of the CUORE-0 tower. Floors 9--11 have lower efficiencies because of a not working detector. Figure from C. Alduino \textit{et al.}~\cite{Q0-120Te}.}
\label{fig:120te_eff}
\end{center}
\end{figure}

At 0.09(1)\% the natural isotopic abundance of \tecw is very low and far from ideal for a rare decay search. However, the emission of a positron followed by a pair of 511-keV $\gamma$ rays provides a very clean experimental signature for the \bpEC decay mode. CUORE and CUORE-0 can make use of the segmented detector array to look for events with different multiplicities. In CUORE-0~\cite{Q0-120Te}, we looked for an energy deposition of 692.8~keV within a crystal, in coincidence with one or two 511-keV $\gamma$ rays in the same or nearby detectors. The possible scenarios for detection are listed in Table~\ref{tab:120te_eff}, and the containment efficiency for each floor is shown in Figure~\ref{fig:120te_eff}. The absorption length of 511-keV $\gamma$ in TeO$_2$ is less than 2~cm and, hence, the probability that it will escape a detector is small. It is, therefore, not surprising that the cleanest of the signature ($\mathcal{M}_3$\xspace) has, unfortunately, the smallest probability of detection for our crystals. The highest probability of detection corresponds to events in which one or both the 511-keV $\gamma$ get absorbed in the primary crystal where the decay takes place. However, in these cases it is difficult to discriminate the background based on the multiplicity of the events. 

\begin{table}[h]
\ttbl{30pc}{Table with signatures for \tecw 0$\nu$-\bpEC decay and its associated multiplicity.}
{\begin{tabular}{lc} \\[6pt]\hline
\hline	Signature &	Multiplicity  	\\
\hline
(0) E$_{692.8}$	& $\mathcal{M}_1$\xspace \\
(1) E$_{692.8+511}$	& $\mathcal{M}_1$\xspace\\
(2) E$_{692.8+511+511}$ & $\mathcal{M}_1$\xspace\\
(3) E$_{692.8}$ + E$_{511}$ & $\mathcal{M}_2$\xspace \\
(4) E$_{692.8+511}$ + E$_{511}$ & $\mathcal{M}_2$\xspace\\
(5) E$_{692.8}$ + E$_{511}$ + E$_{511}$ & $\mathcal{M}_3$\xspace \\
\hline
\end{tabular}} 
\label{tab:120te_eff}
\end{table}

We used the entire CUORE-0 data set, which corresponds to 35.2 kg$\cdot$yr of TeO$_2$ exposure, for the search of \bpEC decay of \tecw. We implemented a simultaneous unbinned Bayesian fit of the energy spectra for all signatures using the Bayesian Analysis Toolkit (BAT) software package. We performed the fit separately on each signature, as well as on all of them together. In the likelihood, we parameterized the signal as an $N$-dimensional Gaussian distribution centered at the characteristic energies reported in Table~\ref{tab:120te_eff}. The background in the ROI was modeled as the sum of a linear background together with a Gaussian for each of the $\gamma$ peaks in the vicinity of the region of interest. The background arising from Compton-scattered $\gamma$ rays, namely from $^{60}$Co and $^{214}$Bi, was also taken into account for the double ($\mathcal{M}_2$\xspace) and triple-coincidence ($\mathcal{M}_3$\xspace) data. 
\begin{table}[ht]
\ttbl{30pc}{Results of the \tecw 0$\nu$-\bpEC decay analysis on each individual signature. A combined analysis of all signatures together improves the limit.}
{\begin{tabular}{cc} \\ \\[6pt] \hline
\hline Signature &  Limit on $T_{1/2}$~[yr]  \\
\hline\\ [-2ex]
(0) &  $2.5\times10^{19}$ \\
(1) &  $1.4\times10^{20}$ \\
(2) &  $4.2\times10^{20}$ \\
(3) &  $4.4\times10^{20}$ \\
(4) &  $1.1\times10^{21}$ \\
(5) &  $1.5\times10^{20}$ \\
\hline\\ [-2ex]
All & $1.6\times10^{21}$ \\
\hline\\ 
\end{tabular}}
\label{tab:120Te_results}
\end{table}
Table~\ref{tab:120Te_results} shows the limit on $T_{1/2}$ for each of the signatures. The efficiencies, the $Q$ value and the isotopic abundance of \tecw were treated as nuisance parameters for the Bayesian fit. We deliberately treated the uncertainty on the energy resolution as a systematic effect to keep the analysis simple. If we varied the energy resolution by $\pm1\sigma$, the limit on the $T_{1/2}$ only changed by $\mp$7\%.  

As can be seen from Table~\ref{tab:120Te_results}, the fit on all signatures provides a limit of $T_{1/2} > 1.6 \times 10^{21}$~yr at 90\% CI for 0$\nu$-\bpEC decay of \tecw. A much stronger limit of $T_{1/2} > 2.7 \times 10^{21}$~yr (90\% C.L.) is obtained by combining the results of CUORE-0 and Cuoricino. This result has a systematic uncertainty of 5\% which is due to the uncertainty on the CUORE-0 energy resolution. 

CUORE has the potential to improve this limit significantly. Apart from having a mass which is about 19 times that of CUORE-0, it is also a closely packed array which will result in an enhancement of efficiencies for signatures with M$>$2. We envisage that we should be able to improve the present limit by at-least two orders of magnitude with CUORE.    

\subsection{$^{128}$Te decays}\label{sec:128te}
$^{128}$Te is the second most abundant natural occurring tellurium isotope, with an isotopic abundance of $\sim$31.7 \%. Like $^{130}$Te, it is a double beta decay emitter, although the decay has never been directly observed. The ratio between its half life and the heavier isotope one has been precisely measured by geochemical experiments to be $\sim 3\times 10^{-5}$, a result determined by the much smaller phase space\cite{Barabash}.

The small Q-value of the decay (865.87 keV\cite{Scielzo:2009nh}) makes its direct detection particularly challenging, both for the already cited small phase space and correspondingly long lifetime, ($2.0 \pm 0.3 \times 10^{24}$ yr\cite{Barabash}), and because the electron energy spectrum is located in a region where the background generated by the environmental $\gamma$ radioactivity is typically quite high. Although a positive observation is beyond the reach of present generation experiments, CUORE will certainly be able to largely improve the direct detection limits both for the two neutrino and neutrino-less decay modes. The MiDBD experiment set a lower limit achieved to the \bbz half-life of this isotope of 1.1$\times$10$^{23}$\,yr~\cite{Arnaboldi_2002}, CUORE, thanks to a more than two orders of magnitude increase in the isotope mass and the reduced background, will be able to achieve a much better sensitivity in a short time.   

\section{Other rare decays}\label{sec:raredec}

Thanks to the performance of the CUORE detector, both in terms of active mass, almost 100\% detection efficiency in a wide energy range and the almost unitary energy quenching (i.e. all particles have a similar energy response despite their different specific ionization), a number of rare decays can be studied with high sensitivity and small systematic effects. These processes include both the decay of nuclei that are currently considered stable, and processes not allowed by the Standard Model whose observation would be a hint of Physics beyond it.

\subsection{Electron Decay}\label{electron decay}
Electron decay is an electric charge-violating phenomenon. The search for charge violating processes is a sensitive test of the Standard Model, as electric charge one of its most relevant building blocks.
Although lepton and baryon numbers are violated in a number of grand unified models, the experimental search is generally limited to decays that violate electric charge while conserving all the other conserved numbers.

Possible final states include a photon and a neutrino, or three neutrinos, in order to conserve lepton number.
\begin{eqnarray}
	e^- \rightarrow \gamma \, \nu_\mathrm{e} \label{gammanu}\\
    e^- \rightarrow \nu_\mathrm{e} \,\, \nu_\mathrm{x} \,\, \bar\nu_\mathrm{x} \label{3nu}
\end{eqnarray}

The search for both final states can profit from CUORE's large detection efficiency, energy resolution, and low threshold.\\
For both the decay modes the X-rays from the de-excitation of the atom whose electron has decayed are expected. Given the very high probability of containing X-rays and Auger electrons in TeO$_{2}$ crystals, an energy deposition corresponding to the binding energy $E_b$ of the decaying electron is expected in the crystal where the decay occurs. In (\ref{gammanu}), an extra energy deposition
\begin{equation}
E_\gamma = \frac{m_e c^2 - E_b}{2}
\end{equation}
is expected when the $\gamma$ ray is absorbed. This can occur in the same crystal where the electron decay took place, or in one of the neighboring crystals, thus allowing a further background reduction by studying the event topology.

\subsection{Lorentz invariance violation}\label{lorentz violation}
Lorentz invariance violation arising from the spontaneous breaking of the underlying space-time symmetry is an interesting theoretical feature that can be parametrized within the so-called Standard Model Extension (SME). In the literature\cite{Diaz:2013ywa}, double beta decay is presented as a process where Lorentz violating effects in the neutrino sector can be investigated. The effect of SME Lorentz invariance violating terms can be studied both in the two-neutrino and in the neutrino-less decay mode.

For what concerns the two-neutrino mode, a distortion of the two-electron summed energy is expected due to an extra term in the phase space factor, with a maximum effect at an energy value that can be predicted by the theory. Therefore an observation would allow to both quantify the magnitude of the violation and confirm the theoretical model describing the physics behind it. For $^{130}$Te, the maximum of the residual spectrum is expected at 1.05 MeV. Between this energy and the Q-value the majority of the events recorded by CUORE are expected to be 
due to the two-neutrino decay (as can be extrapolated from the comparison of backgrounds in Figure~\ref{fig:cuore0+cuore} and signal in Figure~\ref{fig:JAGS-2n}), thus allowing a high precision measurement of the spectral shape and its possible deformation.

The neutrino-less decay mode would also be affected by a Lorentz invariance violating term. Supposing that no Majorana mass term exists, it can be shown\cite{Diaz:2013ywa} that a Lorentz invariant violation can alone generate the process. In this case, the decay rate measurement would be sensitive to the same SME coefficient that controls the CPT violation.

\subsection{$^{123}$Te decay}


The electron capture decay of $^{123}$Te (a naturally occurring isotope of Te) to the ground state of $^{123}$Sb, although expected to occur, has never been experimentally observed. 
This isotope has an isotopic abundance of 0.908$\pm$0.002 and a Q-value of 51.9~keV. Because of the large spin difference between the parent and daughter nuclei, the EC decay occurs primarily from the L and M shells\cite{Bianchetti,TableIsotopes} while the capture from the K shell is suppressed by about 3 orders of magnitude.

A sensitive search for the K-shell EC performed with TeO$_2$ thermal detectors have been able to put only  a lower limit on the half-live of 9.2$\times$10$^{16}$\,yr\cite{Alessandrello:2002ag}. Given the enormously increased exposure (the exposure used in the analysis of the cited result corresponds to a single hour of CUORE data taking), CUORE can improve this result by orders of magnitude.

The advantages of the thermal detector technique are the same already discussed for the electron decay: as the source is part of the active volume, a signal corresponding to the total binding energy of the captured electron can be measured with almost 100\% efficiency.
For the (suppressed) K-capture a line at 30.5~keV is expected, while for the most probable L$_3$ capture (L$_1$ and L$_2$ intensities are about one and three orders of magnitude lower than L$_3$\cite{TableIsotopes}) the line should appear at 4.1~keV. Finally, the M transitions are all beyond any possible detection capability given their too small energy release (of the order of eV). The excellent energy resolution helps in disentangling these low energy signals from the most relevant backgrounds, i.e. the X-ray/Auger electron cascades generated when a hole is left in a $^{123}$Te atom by the interaction of an externally generated $\gamma$. Moreover, in a high granularity array like CUORE, these events have a high chance of generating energy depositions in multiple crystals, while a genuine EC process is most likely a single-crystal event.
Other possible backgrounds can come from the EC of $^{121}$Te and $^{121m}$Te, both generated via neutron activation of the naturally occurring $^{120}$Te. The neutron flux in the underground location of CUORE, however, is small enough to assume the {\it in situ} activation to be negligible, while the activity generated between the growth of the crystals and the storage underground is expected to have already decayed by many orders of magnitude (and could anyway be tagged thanks to its well known time-dependence).

The experimental measurement of $^{123}$Te half life has important theoretical implications: very low rates are predicted if a strong (up to six orders of  magnitude\cite{PhysRevC.56.R1675}) suppression of the nuclear matrix element is generated by the cancellation between particle-particle and particle-hole correlations. An experimental study of this effect would be a severe test of the nuclear models that are used to calculate the matrix elements for rare electroweak decays\cite{Civitarese:2001kh}.


%


\section{Other rare processes}\label{sec:rarepro}
Despite being mainly a tool for studying rare nuclear decays happening inside its active mass, some features of CUORE make it a sensitive probe for the study of other rare processes. Most of these processes have a signature at the lower end of the energy spectrum. The excellent energy resolution of single-particle thermal detectors typically translates into a low energy threshold that, combined with an almost 100\% detection efficiency and signal quenching and low background level make them versatile instruments for searches at low energy as well.

\subsection{Dark Matter search}
As discussed extensively in a dedicated paper\cite{Alduino:2017xpk}, CUORE is characterized by an excellent sensitivity for dark matter searches. In particular, very promising is the sensitivity to WIMP expected seasonal modulation generated by the revolution of the Earth.
The signature of a WIMP interaction is the recoil of a scattered nucleus in the crystal lattice. The recoil energy spectrum is quasi-exponentially shaped and extends to only few tens of keV for a typical WIMP mass of $\sim$100 GeV/c$^2$. As a consequence of the modulating relative velocity of the Earth with respect to the posited WIMPs galactic halo, 
a change of the recoil spectral shape and a corresponding modulation of the interaction rate is expected as a function of time, with well defined frequency and phase. This modulation results in a time-dependent observed rate of events with energy close to the detector energy threshold.

Critical aspects of the detector performance to make it suitable for WIMPs search are, obviously, a low energy threshold\cite{Alessandria:2012le} and large nuclear recoil quenching factor (almost unitary in the case of TeO$_2$ thermal detectors).

The CUORE-0 detector was used as a test bench for the algorithms needed to lower the threshold and for the analysis procedures. A critical aspect of this analysis is the stability of the background count rate, that is strongly related to the stability of the efficiency of the cuts used to select particle events from any spurious event generating a triggered waveform with similar energy. These cuts are mainly based on the shape of the waveform, thus a large effort has been put into the study of their efficiency and the systematic effect associated to its evaluation.

The large nuclear recoil quenching factor is one of the key advantage of thermal detectors when compared to other detection techniques. CUORE-0 data have been used to evaluate this quenching factor by looking for simultaneous energy depositions in facing crystals, where the sum energy equals the Q-value of an alpha decay. A fraction of these events are generated by alpha-decaying contaminations located on the very surface of a crystal, so that the recoiling nucleus is fully contained in one crystal, while the emitted alpha particle is absorbed in the facing one.
This analysis resulted in a value of the quenching factor compatible with unity.

Another aspect of the detector response that has been studied is the energy scale uncertainty at the lower end of the spectrum. Calibration curves are typically built by measuring the detector response to $\gamma$ radiation at energies starting from above 200 keV. The extrapolation of these response functions at lower energy can be checked by measuring the position of peaks associated to the absorption of Te X-rays (around 27 and 30 keV). A negligible systematic shift is measured.

Based on the experience accumulated with the analysis of CUORE-0 data, the expected sensitivity of the CUORE experiment can be evaluated, taking into account the increased active mass and exposure time. The result is that the CUORE sensitivity, after 5 years of data-taking, is expected to be enough to fully explore the parameter region where a positive annual modulation signal is claimed\cite{Bernabei:2015lzt}, providing an important independent contribution to the WIMPs search via direct detection with solid-state devices.

\subsection{Axions search}
Axions (or more generally axion-like particles, ALPs) are considered an elegant theoretical solution to the strong CP problem in QCD. They appear as the Nambu-Goldstone bosons of a spontaneously broken U(1)$_{PQ}$ global symmetry of the strong sector, introduced by Peccei and Quinn\cite{Peccei:1977hh, Peccei:1977ur} to explain the observed upper limit of the neutron electric dipole moment, much smaller than the otherwise predicted one.

Light ``invisible'' axions are considered good candidates for Dark Matter. They are expected to couple both with the electromagnetic field and directly to leptons and quarks, therefore a comparatively large production yield should be associated with dense and hot matter, like the sun's core. 

The $^{57}$Fe 14.4 keV nuclear transition is a good candidate process where a monochromatic axion can be emitted in place of the gamma or conversion electron. This nuclear level can be populated in the inner region of the sun via thermal excitation, and $^{57}$Fe is a natural occurring isotope with and isotopic abundance of 2.2\%.
The axions produced in the Sun can subsequently convert (via the so-called inverse Primakoff effect\cite{Li:2015tsa}) into photons in the electric field of a Te nucleus in the CUORE crystals, generating a 14.4 keV photon that is fully absorbed and detected as a monochromatic peak in the low energy region of the spectrum. As the conversion probability depends on the electron charge distribution, a daily modulation in the counting rate is expected, depending on the orientation of the crystallographic axes of TeO$_2$ with respect to the Sun direction.

A study\cite{Li:2016qsn} is have been published where the sensitivity of the CUORE experiment to this process is computed, based on realistic assumptions for the background levels and energy resolution. An interesting contribution to the exclusion of large axion mass regions of the $m_a$ - $g_{a\gamma\gamma} g^{eff}_{aN}$ parameters space is discussed.

\subsection{Supernova neutrinos and neutrino-nucleus coherent elastic scattering}
Core-collapse supernovae emit a large fraction of the explosion energy through neutrinos. These neutrino fluxes carry rich information on the particle nature (mass, oscillations) and the supernova mechanism and their detection constitutes the only available prompt signal of the supernova explosion itself.

Some present water Cherenkov or liquid scintillator based experiments (Super-Kamiokande \cite{Fukuda:1998fd}, Borexino \cite{Alimonti:2000xc} and LVD \cite{Aglietta:1992dy}) are able to detect supernova neutrinos through Charged Current and Neutral Current scattering on electrons or inverse beta decay processes, but their sensibility is almost limited to electron neutrinos $\nu_{e}$.
An alternative and promising mechanism to detect MeV SN neutrinos is neutrino-nucleus coherent elastic scattering on target nuclei\cite{Amaya:2011sn}. This flavor blind, Neutral Current process presents a highly enhanced cross section for small enough momentum transfer allowing the detection of all neutrinos' flavors ($\nu_{e}$, $\bar\nu_{e}$ and $\nu_{x}$, i.e. the sum of $\nu_{\mu}$, $\bar\nu_{\mu}$, $\nu_{\tau}$, $\bar\nu_{\tau}$) with the same efficiency of a much larger (about 100 times larger) water Cherenkov or liquid scintillator based experiment. The possibility of detecting all neutrinos' flavors gives valuable model independent information on emission spectra. Moreover, neutrino-nucleus coherent scattering,  predicted by the Standard Model, has only recently been observed\cite{Akimov:2017ade}.

The CUORE \cite{Q_Proposal, Q_PRL} experiment can be considered as a  potential ``light" (compared to detectors usually built for this purpose) coherent scattering based supernova observatory. The capability of CUORE of using coherent elastic scattering to detect supernova neutrinos can be studied and its sensitivity calculated based on expected signal and available information on the background and energy threshold.

A supernova neutrino detector must be able to detect a burst of neutrinos with a continuum energy spectrum reaching the Earth in a time interval of few seconds. As long as coherent elastic scattering is the mechanism of interaction in the detector, the signal is given by the recoils of the nuclei inside the detector lattice, hit by the interacting neutrinos. Typical energies for nuclei scattered by supernova neutrinos are below $10$ to $100$ keV (depending on whether the scatter happens on Te or O nuclei), with an exponentially shaped spectrum. Detector threshold is a very important parameter to be taken into account evaluating the detector sensitivity.

Based on \cite{Amaya:2011sn} and assuming a threshold around 3 keV \cite{Alessandria:2012le}), $\sim$ 10-15 neutrino scattering-generated events are expected within the first $\sim$30 keV above threshold for a galactic supernova explosion ($\sim$10 kPc).
These events are exponentially distributed in time with a time constant of $\sim$ 3.5 seconds, and must be compared with the average number of uniformly distributed background events in the same energy window. In CUORE this rate is conservatively expected to be around 1 Hz over the whole detector, making it sensitive to this kind of event.


\section*{Acknowledgments}
The CUORE Collaboration thanks the directors and staff of the
Laboratori Nazionali del Gran Sasso and the technical staff of our
laboratories. This work was supported by the Istituto Nazionale di
Fisica Nucleare (INFN); the National Science
Foundation under Grant Nos. NSF-PHY-0605119, NSF-PHY-0500337,
NSF-PHY-0855314, NSF-PHY-0902171, NSF-PHY-0969852, NSF-PHY-1307204, NSF-PHY-1314881, NSF-PHY-1401832, and NSF-PHY-1404205; the Alfred
P. Sloan Foundation; the University of Wisconsin Foundation; and Yale
University. This material is also based upon work supported  
by the US Department of Energy (DOE) Office of Science under Contract Nos. DE-AC02-05CH11231,
DE-AC52-07NA27344, and DE-SC0012654; and by the DOE Office of Science, Office of Nuclear Physics under Contract Nos. DE-FG02-08ER41551 and DE-FG03-00ER41138.
This research used resources of the National Energy Research Scientific Computing Center (NERSC).
\bibliographystyle{spphys}       
\bibliography{main}

\end{document}